\pdfoutput=1
\RequirePackage{fix-cm}

\documentclass[twocolumn]{svjour3}
\smartqed

\usepackage{amsmath}
\usepackage{graphicx}
\usepackage{epstopdf}
\usepackage{lineno}
\usepackage{hyperref}
\usepackage{xspace}
\newcommand{\pt}{\ensuremath{p_{\mathrm{T}}}\xspace}
\usepackage{cite}
\usepackage{todonotes}

\begin{document}

\title{Particle-based Fast Jet Simulation at the LHC with Variational Autoencoders}

\author{Mary Touranakou \and
        Nadezda Chernyavskaya \and
        Javier Duarte \and
        Dimitrios Gunopulos \and
        Raghav~Kansal \and
        Breno Orzari \and
        Maurizio Pierini \and
        Thiago~Tomei \and
        Jean-Roch~Vlimant
}

\institute{ Mary Touranakou 
            \at European Organization for Nuclear Research (CERN), CH-1211 Geneva 23, Switzerland 
            \at National and Kapodistrian University of Athens, Athens 157 72, Greece
            \and
            Nadezda Chernyavskaya
            \at European Organization for Nuclear Research (CERN), CH-1211 Geneva 23, Switzerland 
            \and
            Javier Duarte 
            \at University of California San Diego, La Jolla, CA 92093, USA
            \and
             Dimitrios Gunopulos
             \at National and Kapodistrian University of Athens, Athens 157 72, Greece
             \and 
             Raghav~Kansal 
            \at University of California San Diego, La Jolla, CA 92093, USA
            \and
            Breno Orzari
            \at Universidade Estadual Paulista, S\~{a}o Paulo/SP - CEP 01049-010, Brazil
            \and
            Maurizio Pierini
            \at European Organization for Nuclear Research (CERN), CH-1211 Geneva 23, Switzerland
            \and
            Thiago~Tomei
            \at Universidade Estadual Paulista, S\~{a}o Paulo/SP - CEP 01049-010, Brazil
            \and
            Jean-Roch~Vlimant
            \at California Institute of Technology, Pasadena, CA 91125, USA
}

\date{Received: date / Accepted: date}

\maketitle
\begin{abstract}

We study how to use Deep Variational Autoencoders for a fast simulation of jets of particles at the LHC. We represent jets as a list of constituents, characterized by their momenta. Starting from a simulation of the jet before detector effects, we train a Deep Variational Autoencoder to return the corresponding list of constituents after detection. Doing so, we bypass both the time-consuming detector simulation and the collision reconstruction steps of a traditional processing chain, speeding up significantly the events generation workflow. Through model optimization and hyperparameter tuning, we achieve state-of-the-art precision on the jet four-momentum, while providing an accurate description of the constituents momenta, and an inference time comparable to that of a rule-based fast simulation.

\end{abstract}

\section{Introduction}

At particle colliders, collimated sprays of particles are produced as a consequence of the parton shower and hadronization processes typical of Quantum Chromo Dynamics (QCD). 
These sprays of particles, called {\it jets}, are reconstructed applying a recombination clustering algorithm, exploiting physics-inspired metrics such as the anti-$k_t$ distance~\cite{antikt}. Often, jets are clustered from energy deposits recorded in the electromagnetic and hadronic calorimeters of a particle detector. 
At the LHC, jets can be clustered from a list of reconstructed particles, the so-called particle-flow (PF) candidates~\cite{PFcms,PFatlas}. 
In this case, jets would be sparse sets of objects (the constituents), each represented by its momentum\footnote{As common for collider physics, we use a Cartesian coordinate system with the $z$ axis oriented along the beam axis, the $x$ axis on the horizontal plane, and the $y$ axis oriented upward. The $x$ and $y$ axes define the transverse plane, while the $z$ axis identifies the longitudinal direction. The azimuth angle $\phi$ is computed with respect to the $x$ axis. 
The polar angle $\theta$ is used to compute the pseudorapidity $\eta = -\log(\tan(\theta/2))$. 
The transverse momentum ($p_T$) is the projection of the particle momentum on the ($x$, $y$) plane. 
We fix units such that $c=\hbar=1$.} and possibly a set of auxiliary features, such as the nature of the particle (electron, muon, etc.), its electromagnetic charge, etc. 

A time- and resource-effective strategy to simulate jet production is a fundamental asset for physics studies at the CERN Large Hadron Collider (LHC). Whether testing predictions of the Standard Model (SM), searching for evidence of physics beyond the SM, or assessing systematic uncertainties associated to a given measurement, physicists rely on an accurate simulation of the full collision process and of the detector response. This implies the need for a detailed simulation of a chain of very different steps, from the proton collision to the generation of the collected signal in the detector sensors. Typically, physicists simulate datasets at least 10 times larger than the amount of collected data, so that the precision on the final measurement is not limited by the amount of simulated data at hand.

A typical high-energy physics (HEP) simulation software relies on the \textsc{GEANT4}~\cite{Agostinelli:2002hh} library to model the interaction of particles traversing the detector material. 
This approach, based on Monte Carlo (MC) techniques, provides a typical accuracy at the percentage level, but it comes at a high cost in terms of computing resource utilization. 
At the LHC, the simulation workflow consumes up to $\sim 50$\% of the total computing resources of an experiment. 
With the amount of collected data increasing, the need for MC simulation is going beyond what the available computing infrastructure could sustain. 
Projected to the planned High-Luminosity LHC upgrade, this trend will eventually become unsustainable~\cite{Alves:2017she}. Jet simulation is one of the most expensive tasks, since jets are very abundant in LHC collisions and are made of many particles. Since the simulation of each of these particles is a demanding operation, the possibility of simulating all particles in a jet at once would be a major improvement. The main difficulty of this task is to match state-of-the-art accuracy, which also depends on the specific use case, e.g., which quantity a specific analysis uses. For instance, an algorithm reproducing the jet kinematic but not describing the angular distribution of the particles in the jet might be suitable for an analysis needing a good description of the jet momentum (e.g., a dijet resonance search), but not for an analysis exploiting jet substructure techniques (e.g., an all-jet diboson resonance search). Certainly, an algorithm describing every aspect of jet physics would be an ideal solution.

As a first step, a typical LHC simulation makes use of an {\it event generator}, modeling a proton-proton collision, the consequent production of quarks and gluons (among other particles), and their hadronization into jets of particles. Since no detector response is involved, this step typically requires a relatively modest amount of computing resources~\footnote{This picture could in principle change if next-to-leading order precision would be adopted as a default. On the other hand, ongoing work on parallelizing event generation libraries on GPUs~\cite{Hagiwara_2013} may compensate for this precision increase.}. In addition, its content is independent of experimental aspects (reconstruction software, detector configuration, etc.), so that a dataset of these {\it generator-level} events can be stored for long term and used many times (as is the case for the CMS experiment). 
Computing requirements significantly increase when detector effects are to be taken into account. At first, one typically uses \textsc{GEANT4} to simulate the detector response. Then, the reconstruction software runs on the event and produces the objects (e.g., the PF candidates for CMS), eventually clustered into jets. These two steps could be bypassed by a {\it jet response function}, taking as input the list of jet constituents at generator level and returning the list of constituents at reconstruction level. In this paper, we aim at approximating this jet response function with a Variational Autoencoder (VAE), trained using generator-level jets as input and the corresponding reconstruction-level jets as a target. We represent a jet as a list
of particles' momenta. Doing so, the VAE returns a reconstructed jet in a format that is already compatible with a typical PF-based  analysis software. A different approach to the problem of data sparsity consists of representing the jet as a point cloud, as proposed for many HEP-specific problems~\cite{Shlomi:2020gdn}. We investigated that approach when training a Generative Adversarial Network~\cite{kansal2021graph} to generate the list jet constituent momenta from random numbers. A similar approach was presented in Ref.~\cite{hariri2021graph}, where a graph VAE was used to generate detector {\it hits} in a jet, from which jet constituents could be reconstructed using standard rule-based algorithms, e.g., PF reconstruction~\cite{PFcms,PFatlas}. This work has many common points with Ref.~\cite{hariri2021graph}, with two main differences: we do not use graph architectures, and we aim at learning the detector response and bypassing the standard rule-based reconstruction algorithms (to offer further speed up of the simulation process). We do so by taking the reconstructed jet as a target. In this respect, our algorithm could be used to replace detector parametrization approaches now used in Fast Simulation tools~\cite{deFavereau:2013fsa,sekmen2017recent,atlascollaboration2021atlfast3}, while the algorithm of Ref.~\cite{hariri2021graph} aims at speeding up a GEANT-based full simulation. In the future, both approaches will be useful to HEP experiment and, most likely, the ultimate generative model will emerge from a combination of the two. 

This paper is organized as follows: Section~\ref{sec:related} discusses related work. Sections~\ref{sec:data}~and~\ref{sec:arch} describe the benchmark dataset and the model architecture, respectively. A strategy to apply the model to realistic use case is discussed in Section~\ref{sec:application}. Training results are discussed in Section~\ref{sec:results}. Conclusions are given in Section~\ref{sec:conclusions}.

\section{Related Works}
\label{sec:related}

In the recent past, several studies explored the possibility of speeding up the data simulation process using generative models based on deep neural networks (NNs). 
In particular, convolutional neural networks (CNNs) have been proposed to generate single-particle showers in a calorimeter~\cite{Paganini:2017hrr,Paganini:2017dwg,Erdmann:2018jxd,Salamani:2645142,Belayneh:2019vyx,Buhmann:2020pmy,Buhmann:2021vlp}, full jets at the LHC~\cite{deOliveira:2017pjk,Musella:2018rdi,Carrazza:2019cnt}, multi-dimensional functions of kinematic quantities~\cite{Otten:2019hhl,Hashemi:2019fkn}, event kinematics at colliders~\cite{DiSipio:2019imz,Butter:2019cae}, and cosmic ray showers~\cite{Erdmann:2018kuh}. 
Both generative adversarial networks (GANs)~\cite{Goodfellow:2014upx,arjovsky2017wasserstein,1704.00028} and variational autoencoders (VAEs)~\cite{kingma2014auto} were considered. 

These studies clearly demonstrate that integrating deep generative models in the data simulation workflows of HEP experiments could lead to an important saving in terms of computing resources.  
But there is an objective difficulty when scaling up these proof-of-concept solutions to production-ready simulation tools. 
The main problem lies in the complexity of a typical HEP detector, characterized by detector elements with different technology and geometry, partially overlapping with each other and with passive material (e.g., absorbers in calorimeters) in between. As a consequence of this, a typical HEP dataset consists of a sparse set of energy deposits, which often cannot be represented as a regular grid of pixels. Future detectors will be characterized by higher granularity, with
small-size sensors designed to resolve hits from individual particles in dense environments, such as jet cores. This will make the sparsity of the event even more complicated. 
This is the main reason why most of the great ideas based on CNNs had so far a little impact on HEP experiments. 
Instead, other approaches  were explored as alternatives to CNNs, e.g., a recurrent neural network (RNN) trained adversarially~\cite{Martinez:2019jlu}, graph neural networks~\cite{kansal2021graph,hariri2021graph,Belavin_2020}, or normalizing flows~\cite{Krause:2021ilc}. Similar issues are present in other domains, e.g., galaxy simulation in cosmology~\cite{Lanusse}.

In this paper, we investigate an alternative strategy to overcome difficulties with the peculiar nature of HEP data. In previous studies, we discussed how to sample jets as sparse data from a probability density function, modeled using deep generative models. 
To this purpose, we considered both GANs~\cite{kansal2021graph} and VAEs~\cite{Orzari:2021suh}. Here we take a different approach, in which the input is a generator-level jet (as opposed to a vector of random coordinates in some latent space)  
and the aim of the training is to learn a morphing function from generator to reconstruction level. 

The strategy is similar to what is discussed in Ref.~\cite{Chen:2020uds}, where a similar approach is followed to morph a set of analysis-specific features from generator- to reconstruction-level precision.

\section{Dataset}
\label{sec:data}

The reference dataset consists of jets generated in $pp\to WW$ collisions at a center-of-mass energy $\sqrt{s}= 13$~TeV. 
The $W$ bosons are forced to decay to quarks, that then shower to jets. 
The event generation is performed using {\tt PYTHIA8}~\cite{Sj_strand_2015}. 
The generated list of particles is passed to {\tt DELPHES}~\cite{deFavereau:2013fsa}, which applies detector effects using the CMS {\tt DELPHES} description. At this stage, additional collision events are superimposed to the generated collision, to mimic the effect of so-called {\it pileup}. The number of collisions is randomly sampled from a Poisson distribution with expectation value set to 50, in agreement with the expected LHC running conditions for Run 3. The {\tt DELPHES} particle-flow reconstruction algorithm is applied to the event, returning the list of reconstructed particles. Reconstructed particles are required to have $p_T > 250$~MeV and be within $\lvert \eta \rvert < 3.2$. These particles are then clustered into jets using the anti-$k_t$~\cite{antikt} algorithm with jet-size parameter $R=0.5$. Jets with $\pt>200$~GeV and within $\lvert \eta \rvert <2.5$ are retained. These jets represent the target dataset. With the same setting, generator-level jets are clustered from the stable and detectable particles produced in the collision, before detector effects are taken into account. These jets represent the input dataset. 

Target jets within $p_T$ and $\eta$ acceptance are matched to input jets minimizing the angular distance $\Delta R = \sqrt{\Delta \phi^2 + \Delta \eta^2}$. 
An input-target pair is formed taking the closest input jet to each target jet.
For both input and target jets, constituents are ordered by decreasing $\pt$ and the first 50 particles are retained. When fewer particles are present, the list is zero-padded.
The list contains the momentum of each constituent in Cartesian coordinates $(p_x, p_y, p_z)$. The constituent mass is implicitly assumed to be zero. The main advantages of this specific choice are: it retains information of the position of the jet in the detector (as opposed to local coordinate choice); the distribution of these quantities is unbounded and symmetric around $0$, which makes the learning process easier. 
In particular, we avoid issues related to the periodicity of $\phi$ and hard-threshold at boundaries (e.g., on $\pt$). 

We apply feature-dependent standardization by subtracting the mean and scaling the features to unit variance. 
During early stages of this work, we verified that these choices help the model training to converge to more accurate configurations of the network weights. 
After this pre-processing, each jet is represented as a 2D array of $3\times 50$ numbers. 
The whole dataset includes $\sim 1.7$M jets. We split these data in three parts: 60\% for training, 20\% for validation, and 20\% for testing. The dataset is published on Zenodo~\cite{touranakou_mary_2022_6047873}.

\begin{figure*}[t!]
    \centering
    \includegraphics[width=0.9\textwidth]{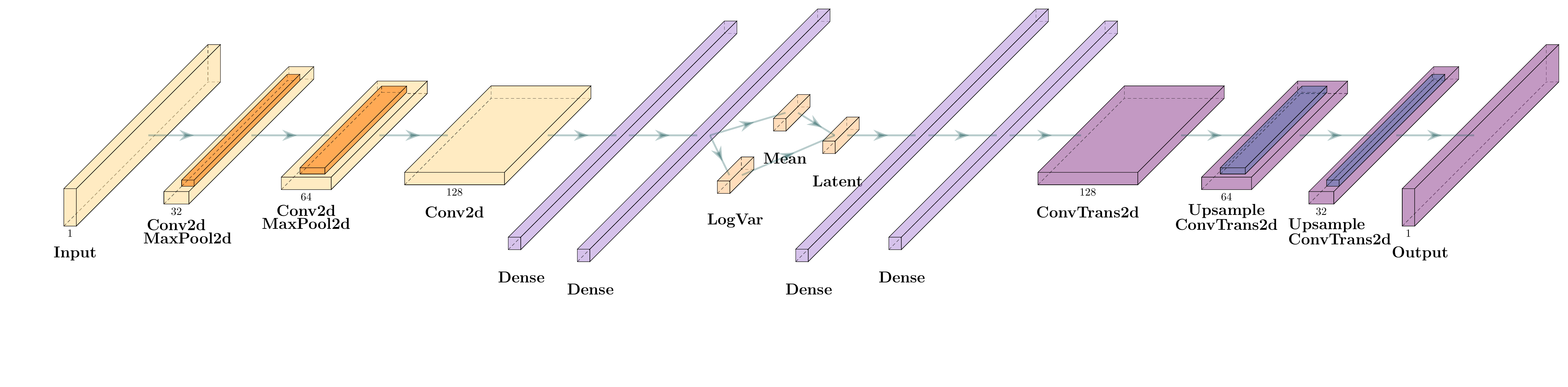}
    \caption{Graphical representation of the VAE architecture.\label{fig:VAE_arch}}
\end{figure*}

\section{VAE Architecture}
\label{sec:arch}

The VAE architecture is schematically shown in Fig.~\ref{fig:VAE_arch}. 
The encoder receives a single-channel $3\times 50$ table, which is processed by three 2D convolutional layers, with 32 3$\times$5 kernels, 64 1$\times$5 kernels, and 128 1$\times$5 kernels, respectively. 
The stride is set to 1 and zero padding is used when the kernel arrives at the edge of the table. 
The output tensor is flattened and passed to two dense layers, with 640 and 150 nodes, respectively.
From the second layer, two 20-dimension vectors are derived, corresponding to the mean $\mu$ and log-variance values of the latent-space variables $z$. 
These values are used to define the Gaussian prior function from which a set of $z$ values is sampled and passed to the decoder. 
The decoder architecture mirrors the encoder, with the Conv2D layers being replaced by ConvTrans2D layers. 
Leaky ReLU activation functions are used across the whole architecture with the coefficient of the negative slope set to 0.1, except for the encoder and decoder output layers, for which a linear activation function is used. Max pooling with the kernel size 1$\times$2 and stride of 2 is applied after each of the first two convolution operations in the encoder. Respectively, two upsampling operations are used in the decoder, each placed before the last two ConvTrans2D layers with a bilinear interpolation scheme. Dropout is added to the output of the first dense layer in the encoder and to the output of the two dense layers in the decoder with a dropout rate of 0.2. The deep learning (DL) model is implemented in PyTorch~\cite{torch}. 

The model is trained using an input dataset, containing the list of particles at generator level, and a target dataset, containing the corresponding list after detector effects and event reconstruction. In this way, the model is trained to regress the detector response function starting from a generator-level jet, i.e. it corresponds to what a Fast Simulation software in HEP computing literature. For this kind of application, a typical state-of-the-art simulation has a 10\% accuracy on jet kinematic properties.

The model training is performed minimizing a domain-specific loss function: 
\begin{align}
    L^\mathrm{VAE} = \frac{1}{N} \sum_{i=0}^N \left [ \beta D^i_\mathrm{KL} + (1-\beta) \left ( L_\mathrm{R}^i + \right . \right. \nonumber \\
    \left . \left . + ~\alpha_m \left (m^i_\mathrm{jet} - \hat{m}^i_\mathrm{jet} \right )^2 + \alpha_{p_\mathrm{T}}
    \left (\pt^{\mathrm{jet},i} - \hat{p}_\mathrm{T}^{\mathrm{jet},i} \right )^2 \right ) \right ]
    \label{eq:loss_penalty}
\end{align}
where $N$ is the dataset size, $L_\mathrm{R}$ is the reconstruction loss (i.e., a distance between the target and the output), $D_\mathrm{KL}$ is the Kullback–Leibler (KL) divergence regularizer usually employed to force the data distribution in the latent space to a multi-dimensional Gaussian with unitary covariance matrix~\cite{rezende2016variational}, and $\beta$ is a parameter that controls the relative importance of the two terms~\cite{Higgins2017betaVAELB}. 
The reconstruction loss $L_\mathrm{R}$ is computed using the permutation-invariant Chamfer loss~\cite{chamfer}:
\begin{equation}
    L_\mathrm{R} = \sum_{i} \min_{j}(p_i-\hat{p}_j)^2 + \sum_{j} \min_{i}(p_i-\hat{p}_j)^2~~,
\label{eq:chamfer}
\end{equation}
where $p_i$ is the feature for the $i$-th target particle, and $\hat{p}_j$ is the corresponding quantity for the $j$-th output particle. 
By construction, 
this quantity is invariant under the permutation of the input or output particle lists. 
We also experimented with a mean squared error (MSE) loss, observing typically worse results. 

In Eq.~(\ref{eq:loss_penalty}), $\pt^\mathrm{jet}$ and $m^\mathrm{jet}$ are the transverse momentum and mass of a target jet, respectively. Jet features are computed from the momenta of the target-jet constituents, while $\hat{p}_\mathrm{T}^\mathrm{jet}$ and $\hat{m}^\mathrm{jet}$ are the corresponding quantities computed from the model output. The coefficients $\alpha_m=1.0$ and $\alpha_{p_\mathrm{T}}=0.1$ were chosen such that the reconstruction loss $L_\mathrm{R}$ and
 the jet-$p_T$ and jet-mass MSE constraints in Eq.~(\ref{eq:loss_penalty}) have similar magnitudes.
The expression in Eq.~(\ref{eq:loss_penalty}) is only one of the possible ways one could enforce kinematic constraints on the jet generator. 
Similar approaches have been followed in previous works, e.g., for particle energy in GAN-based single-particle generators~\cite{Paganini:2017dwg,Belayneh_2020}. 
The main difference here is that the quantity on which the constraint is applied on is analytically computed from the output list, as opposed of being regressed from an image. 
We also tried other combinations of kinematic constraints, e.g., the three momentum components in Cartesian coordinates, observing similar or worse results after training. 
\section{Target application}
\label{sec:application}

The aim of this work is to create a fast-simulation workflow for an analysis demanding a large sample of multijet events. As a reference, we consider the case of dijet resonance searches. Traditionally, these searches are carried on through a bump-hunt maximum-likelihood fit, in which the background is analytically modeled~\cite{CMS:2015xau,ATLAS:2017eqx}. Large samples of simulated multijet events are used to find an adequate model. In addition, a novel simulation-assisted strategy uses ratios of simulated distributions to avoid the need of a specific background analytical model~\cite{CMS:2019gwf}. Also in this case, having at hand a large simulated sample is crucial. Finally, the proposed strategy could be crucial to move the default event-generation precision to next-to-leading order, trading simulation computing time for generation computing time. This would be also relevant for analyses exploiting angular information about the dijet system~\cite{CMS:2018ucw}. Similar considerations hold for multijet searches.

The reference analysis requires an accurate model of jet kinematic properties for jets momenta larger than 200 GeV. As we will see, this can be achieved. But some care is required to model the sharp threshold at 200 GeV. As we experienced in the early stages of this study, an ML-based simulation struggles to model such a sharp threshold. As a solution, we extend the jet phase space in the training sample down to $p_T>130$~GeV~ and we apply the selection of $p_T>200$~GeV on the jets obtained as output of the VAE. A similar problem exists for the $p_T$ threshold of the jet constituents. In this case, we also extend the $p_T$ range of the predicted model down to $p_T>0$~MeV and apply the selection $p_T >250$~MeV afterwards. Similar considerations hold for $\eta$, where the acceptance requirements are imposed on the predicted jets after inference. 

As discussed in Section~\ref{sec:results}, this setup provides an adequate description of the jet kinematic but it fails in providing an accurate description of jet substructure. In this respect, the proposed model cannot be extended to other dijet bump-hunt analyses, e.g., diboson resonance searches, where the accurate modelling of jet substructure is crucial. One could have then enforced a limited scope from the beginning and avoided generating jet constituents, working directly at the level of jet four momenta. However, we see two added values in working with jet constituents: on the one hand, we obtain a faithful description of the jet mass, the most crucial jet-substructure high-level features; on the other hand, we establish a baseline model which could further improve to also model jet substructure. This will be the subject of future studies exploiting a permutation-equivariant graph architecture.

\section{Results}
\label{sec:results}


We train all  models using the Adam~\cite{adam} optimizer with a learning rate of 0.0001 for 300 epochs. The training was repeated for several values of $\beta$, and the value corresponding to the best agreement between input and target ($\beta=1/9$) was chosen. During the training, we monitor the values of the total loss and its individual components evaluated on the training and validation datasets to check for over-training. To quantitatively evaluate the performance of different training settings, in addition to comparing the loss values, we also use the symmetrized  version of the KL divergence (KLD)~\cite{kullback1951information} between probabilities of the predicted and target jet-kinematic distributions (mass, $\pt$, energy, $\eta$, $\phi$). The KLD is computed every 10 epochs on the testing dataset, after rescaling the DL-predicted and target distributions so that the reconstructed distribution is contained in the $[0,1]$ range. The best model is selected based on the values of total and individual components of the loss and the KLD, while ensuring no over-training. 

\begin{figure*}[t!]
    \centering
    \includegraphics[width=0.4\textwidth]{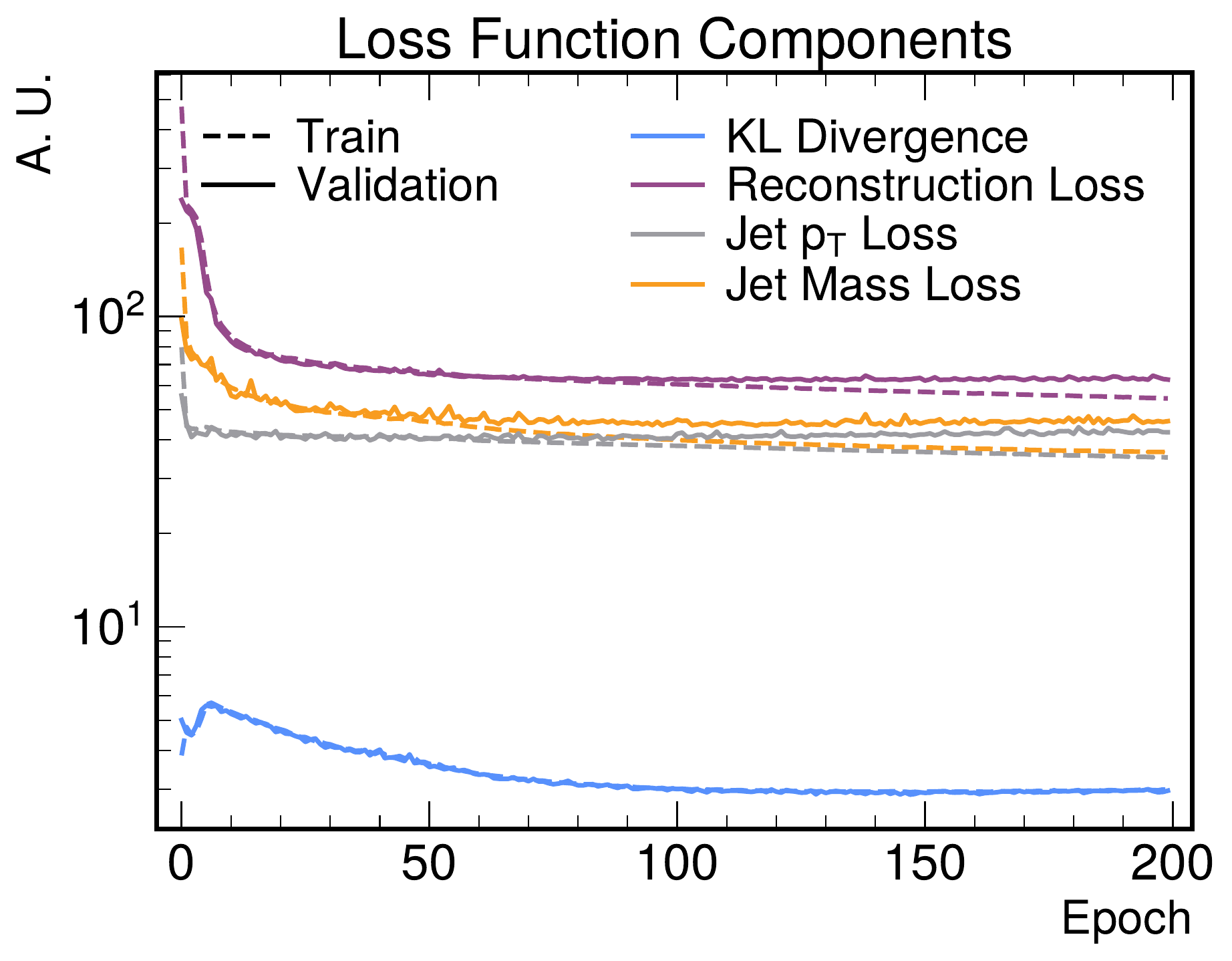}
    \includegraphics[width=0.4\textwidth]{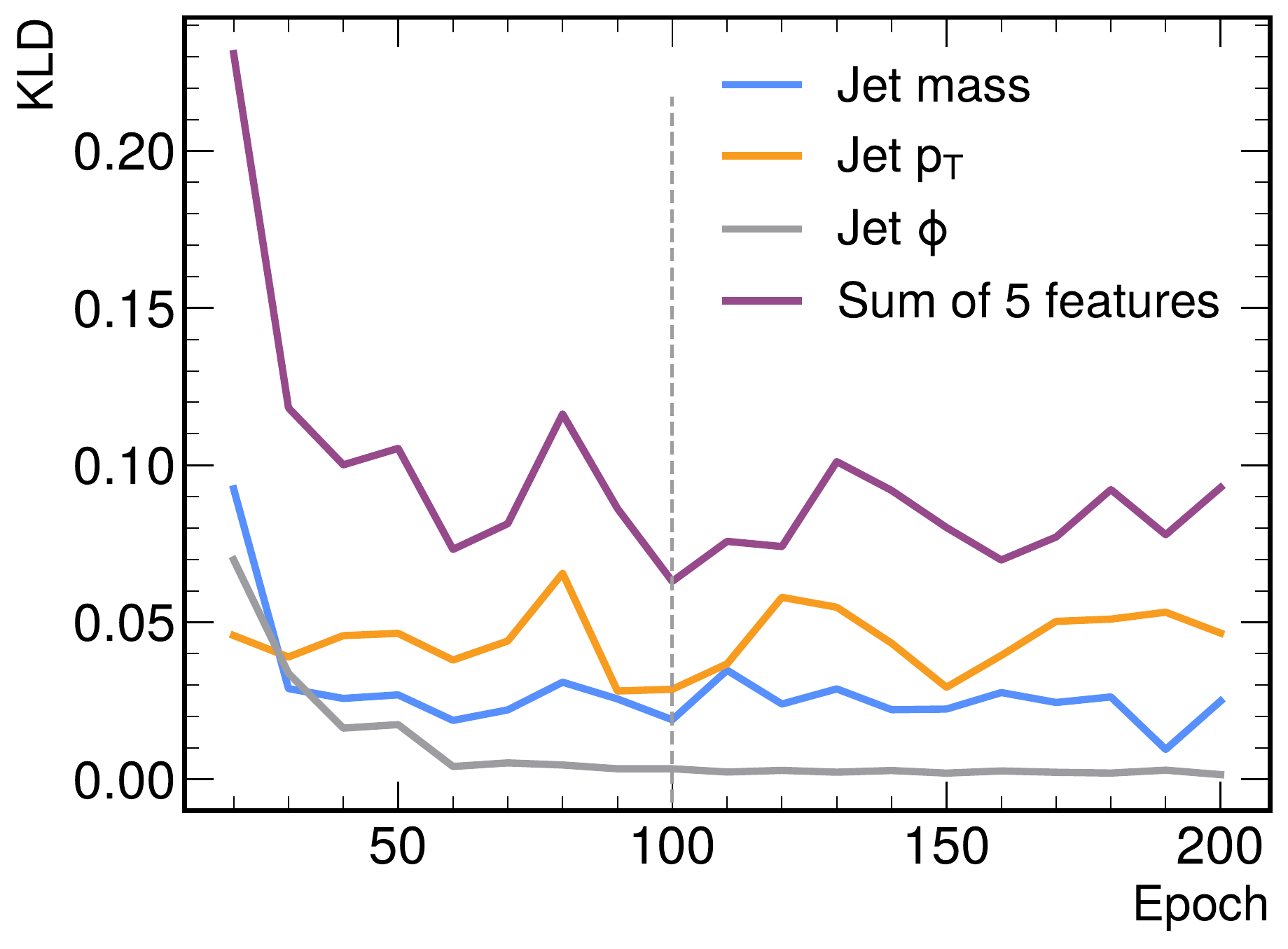}
    \caption{(Left) Evolution of various contributions to the loss, evaluated on the training (dashed lines) and validation (solid lines) datasets, as a function of the training epoch. (Right) Evolution of the KLD as a function of the training epoch. The KLD is computed on the testing dataset. The total KLD, a sum of the KLD for the 5 jet features (mass, $\pt$, energy, $\eta$, $\phi$), and three individual components are shown. The dashed vertical line indicates the chosen epoch. \label{fig:training_loss}}
\end{figure*}

Figure~\ref{fig:training_loss} (left) shows the evolution of various contributions to the loss function (see Eq.(\ref{eq:loss_penalty})), evaluated on the training and validation datasets, as a function of the epoch. The monitored evolution of the KLD computed on the testing dataset for the three jet features (mass, $\pt$, $\phi$) and the total KLD sum of 5 features (mass, $\pt$, energy, $\eta$, $\phi$) is shown in Fig.~\ref{fig:training_loss} (right). The model from epoch number 100 is chosen as the best. 

\begin{figure*}[t!]
    \centering
    \includegraphics[width=0.31\textwidth]{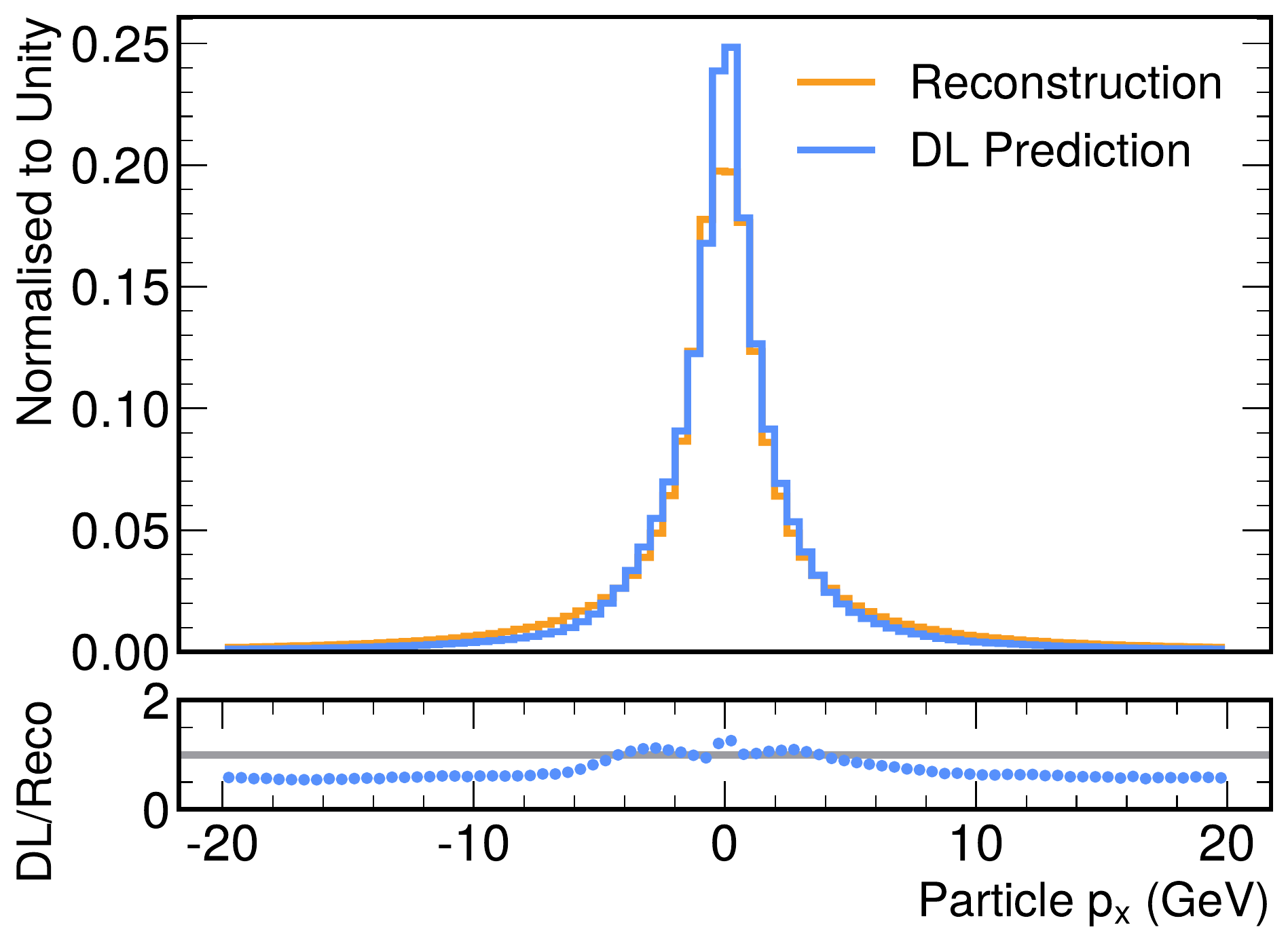} 
    \includegraphics[width=0.31\textwidth]{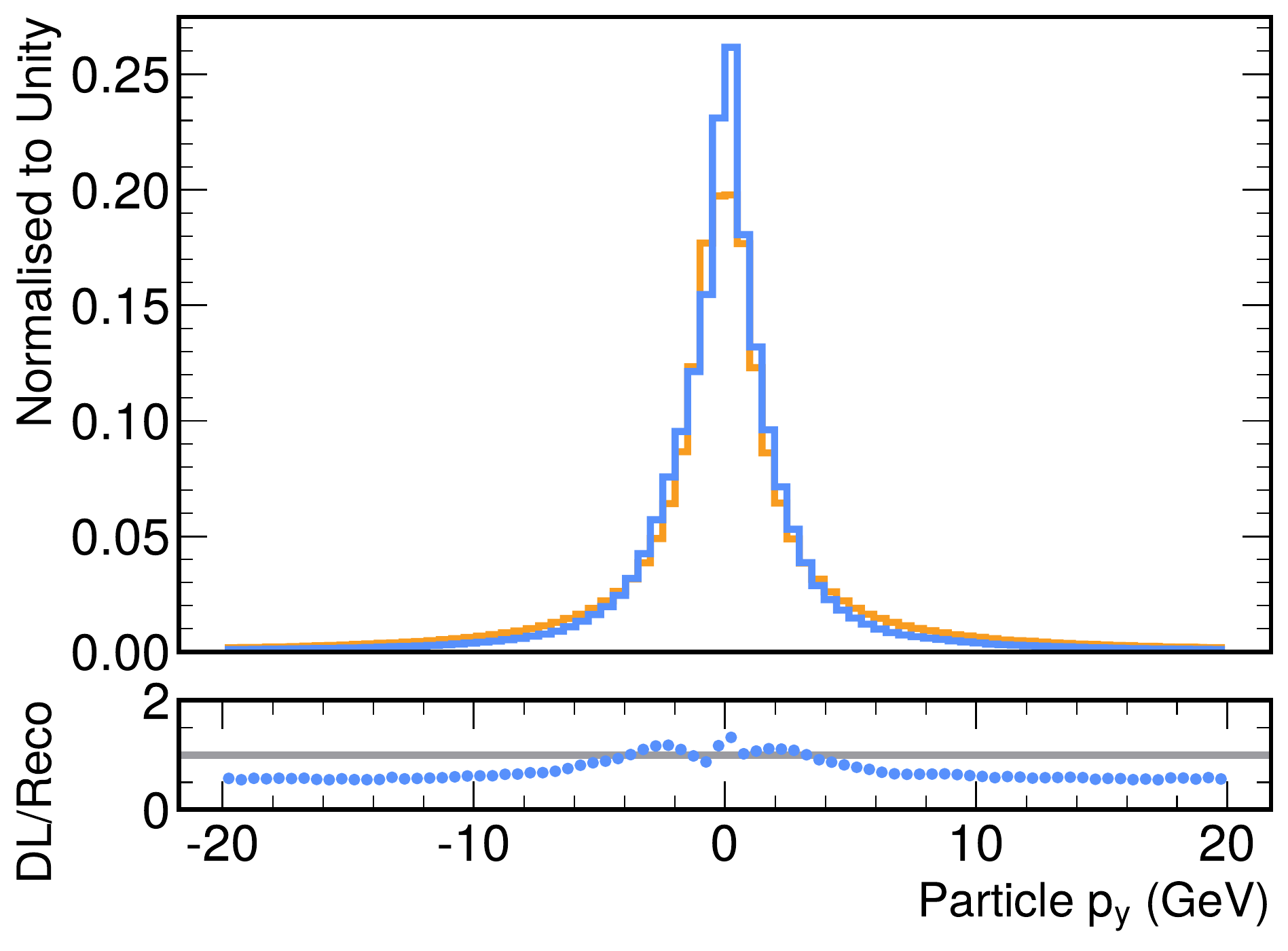} 
    \includegraphics[width=0.31\textwidth]{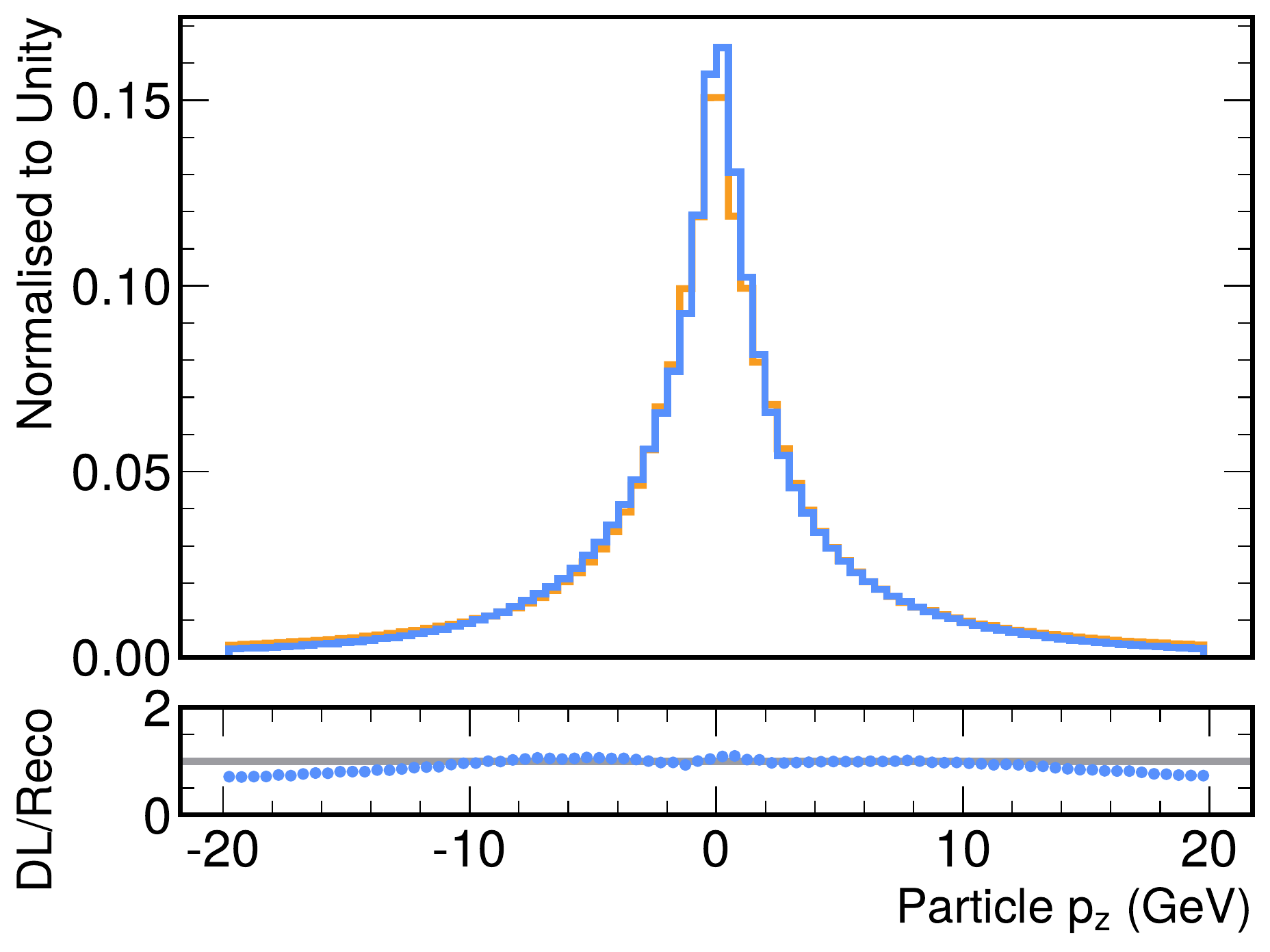} 
    \caption{Comparison of the jet-constituent  $p_x$ (left), $p_y$ (center), and $p_z$ (right) distributions, for the output (DL prediction) and target (Reconstruction) datasets. In the bottom panels, the ratio between the two distributions is shown. These distributions are obtained removing the zero-padding particles from the target list, and enforcing on the output of the DL model the same acceptance requirements that define the jet constituents (see section~\ref{sec:data}). \label{fig:particle_reco}}
\end{figure*}

We compare the distributions of the DL predicted and the target $p_x$, $p_y$, and $p_z$ of the jet constituents in Fig.~\ref{fig:particle_reco}. These distributions are obtained applying the constituents acceptance thresholds (see Section~\ref{sec:data}) to the list of particles which is output by the VAE, as discussed in Section~\ref{sec:application}. We observe a good agreement between the model prediction and the target reconstruction. 

\begin{figure*}[t!]
    \centering
    \includegraphics[width=0.4\textwidth]{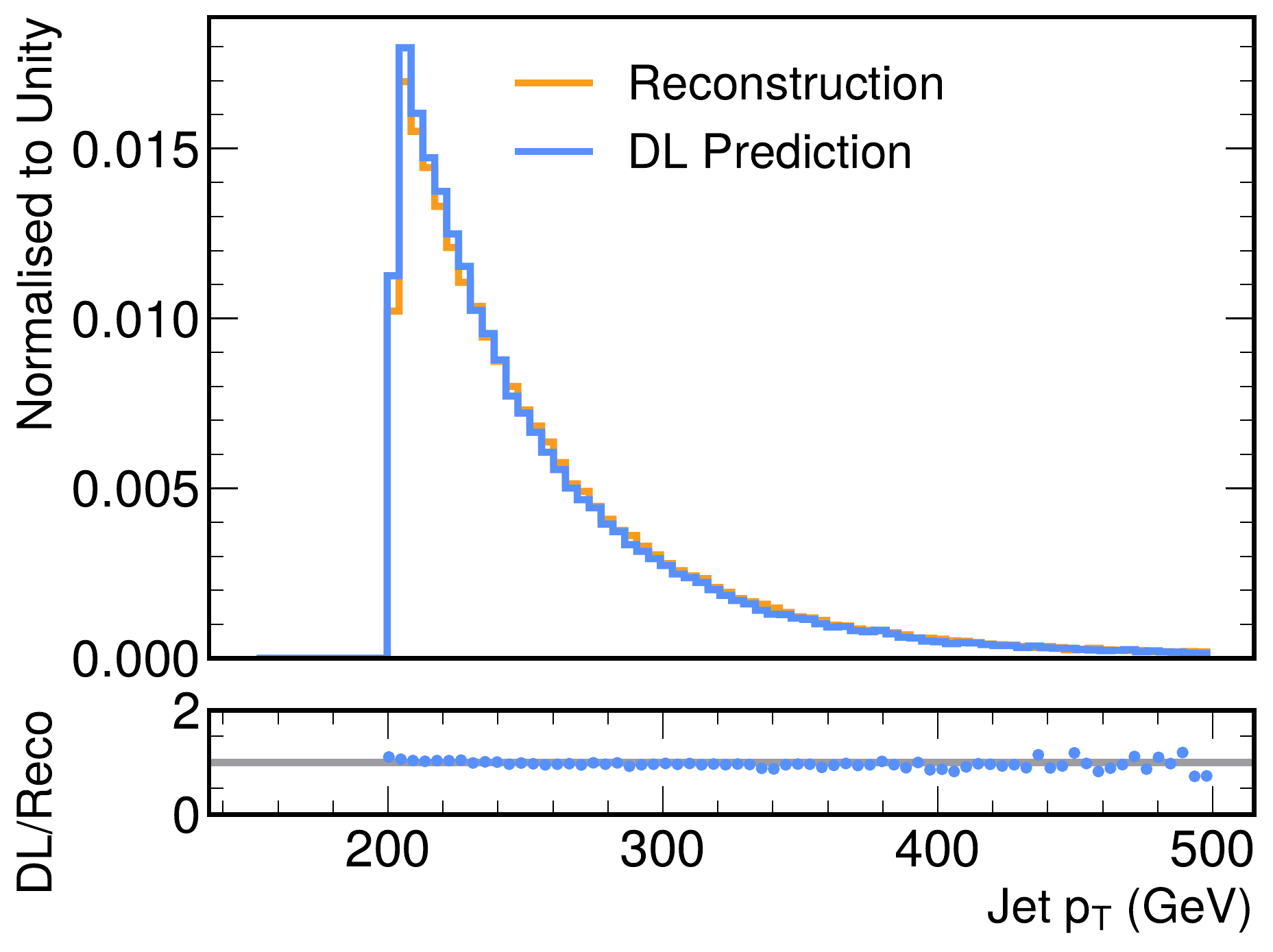}
    \includegraphics[width=0.4\textwidth]{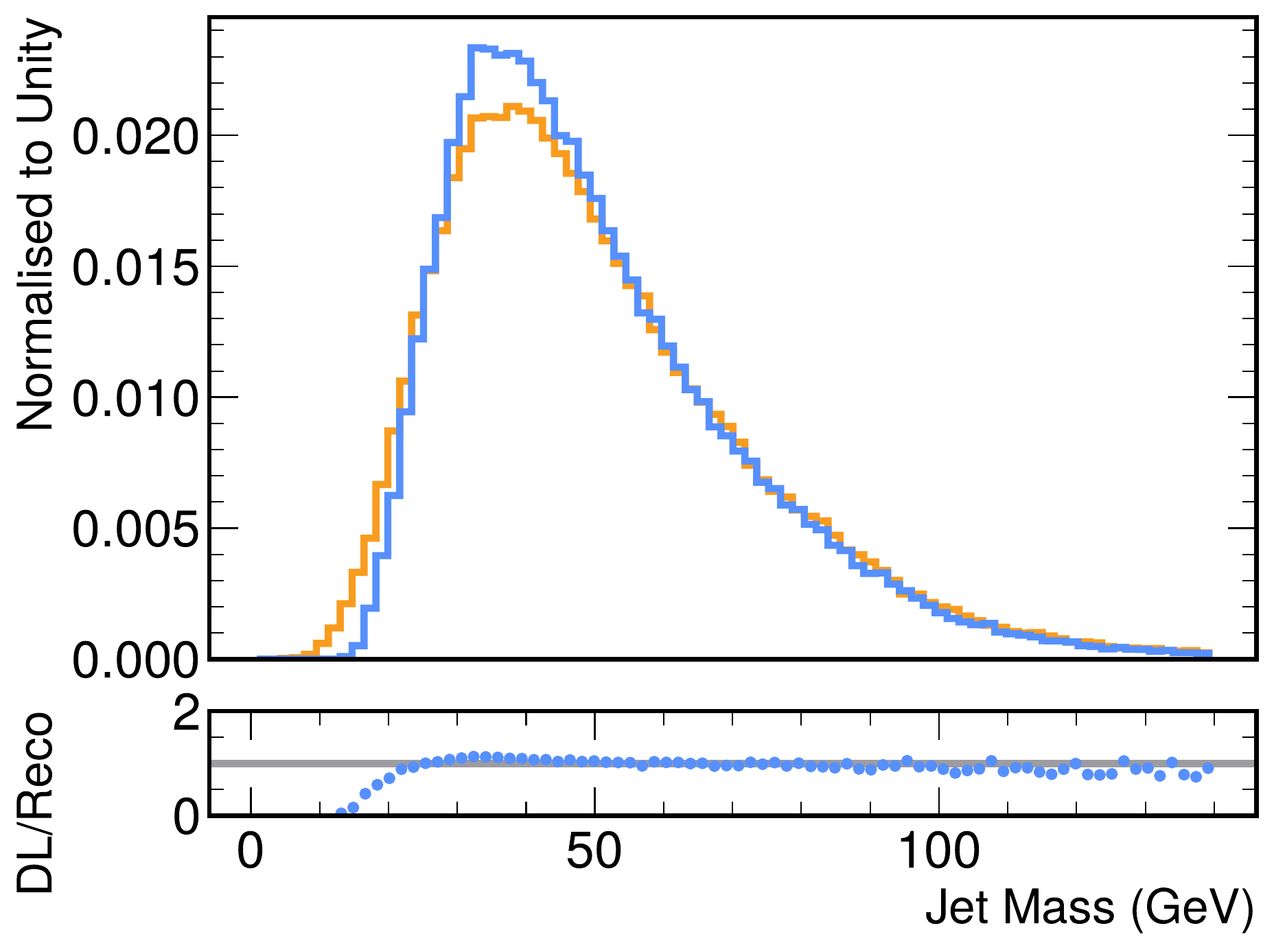}
    \caption{Comparison of the jet $p_T$ (left) and mass (right) distributions, for the output (DL prediction) and target (Reconstruction) datasets. In the bottom panels, the ratio between the two distributions is shown. These distributions are obtained removing the zero-padding particles from the target list, and enforcing on the output of the DL model the same acceptance requirements that define the jet constituents (see section~\ref{sec:data}). \label{fig:jet_reco}} 
\end{figure*}

\begin{figure*}[t!]
    \centering
    \includegraphics[width=0.31\textwidth]{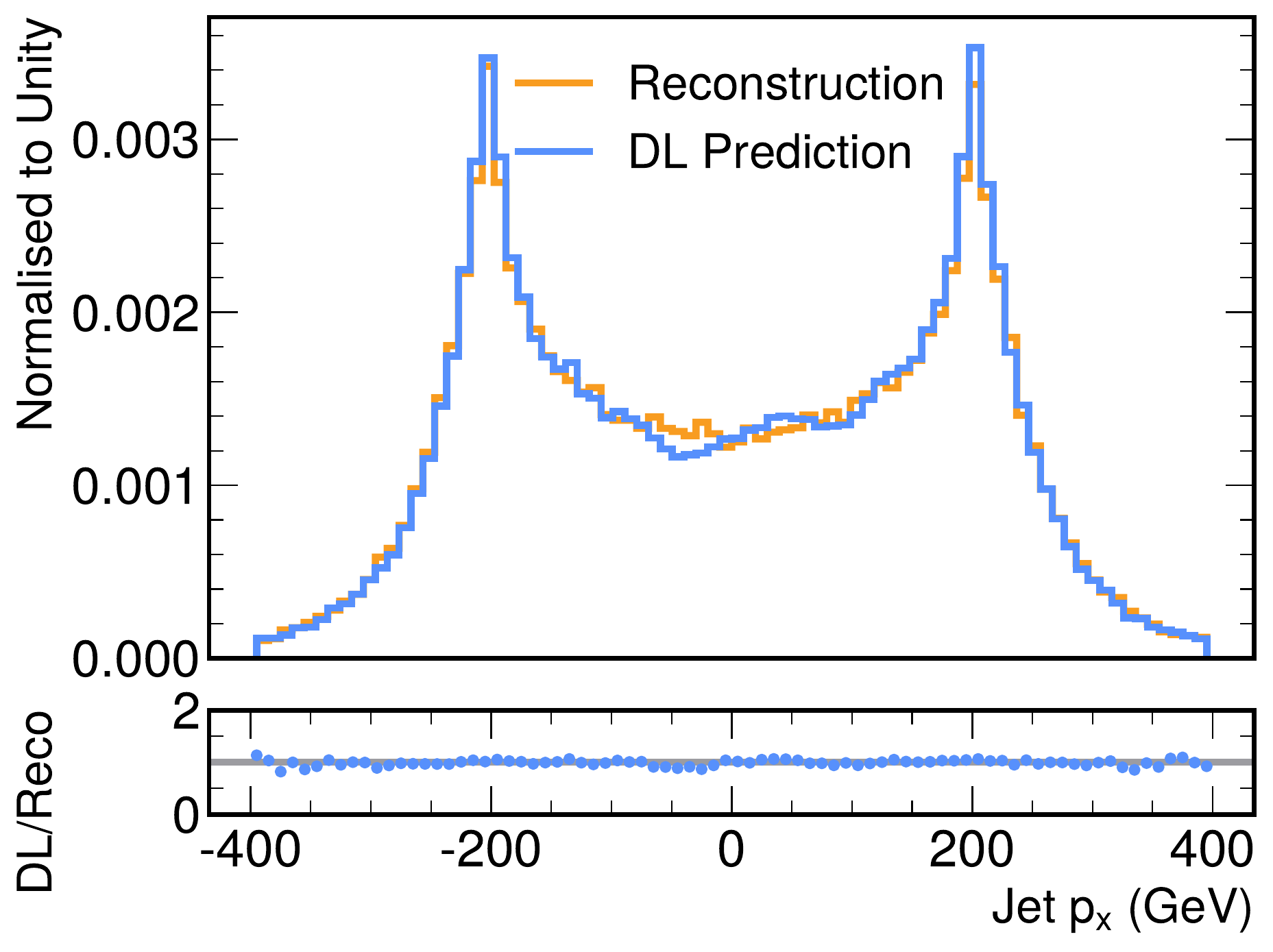} 
    \includegraphics[width=0.31\textwidth]{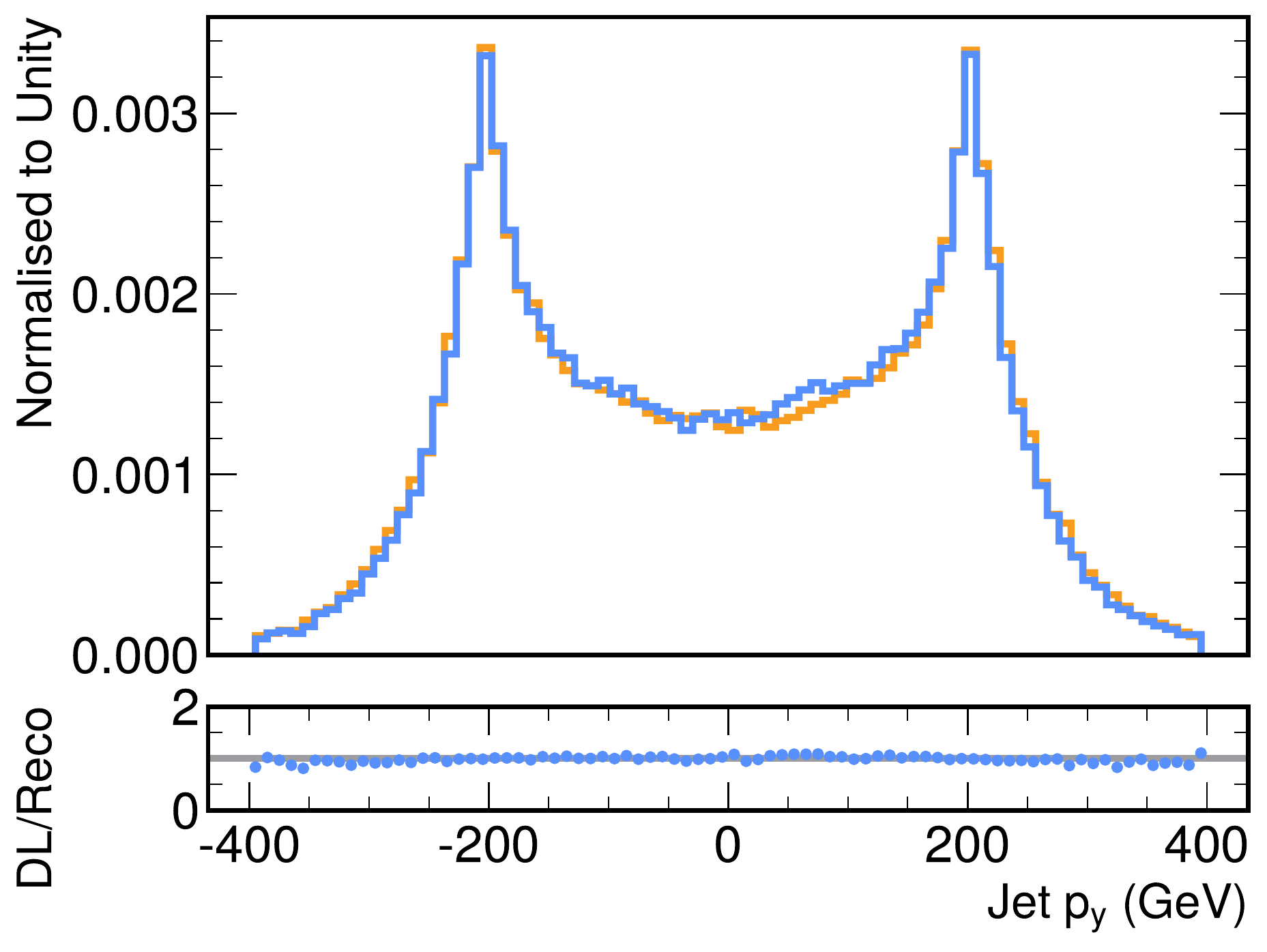} 
    \includegraphics[width=0.31\textwidth]{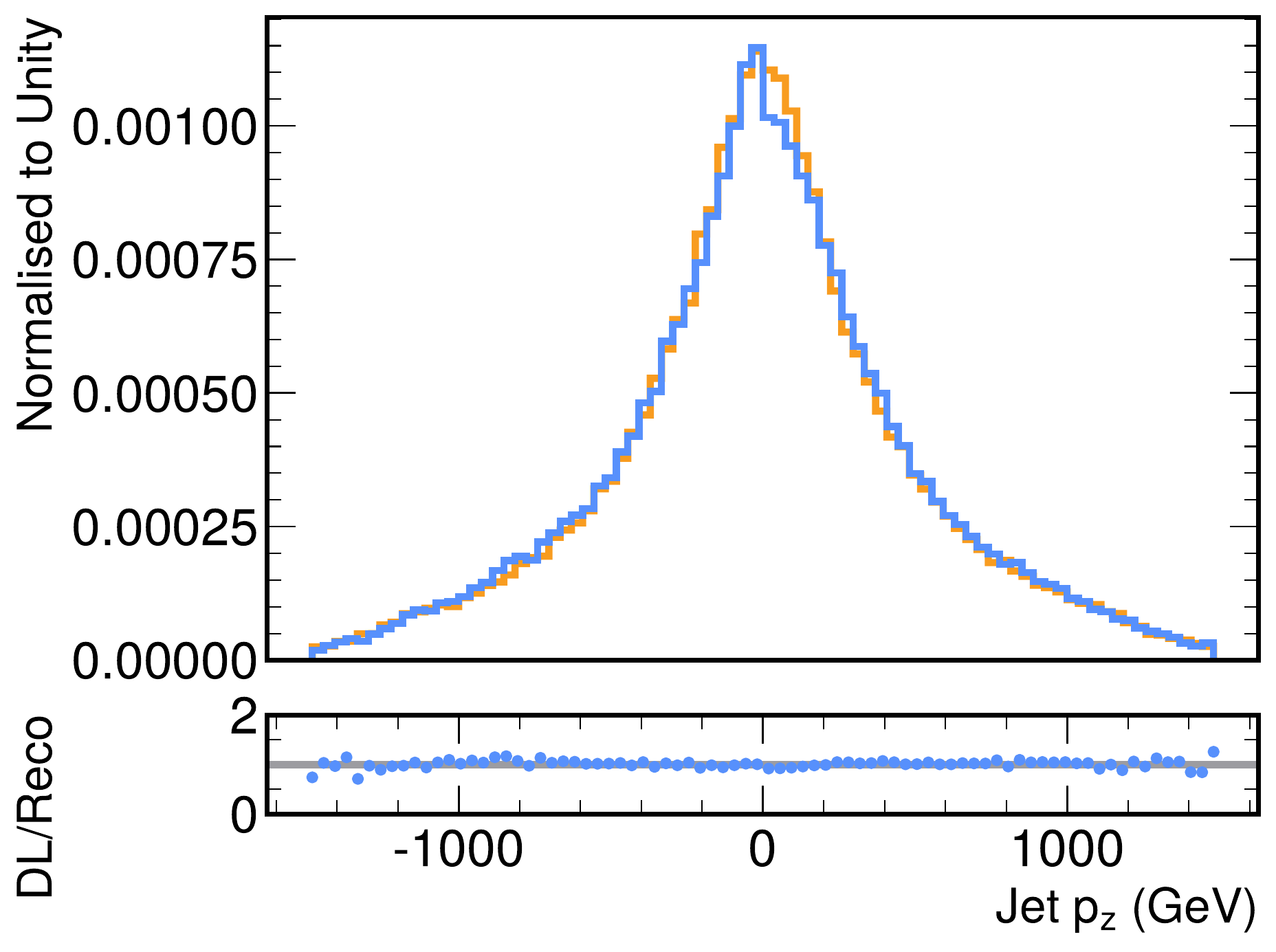} \\
    \includegraphics[width=0.31\textwidth]{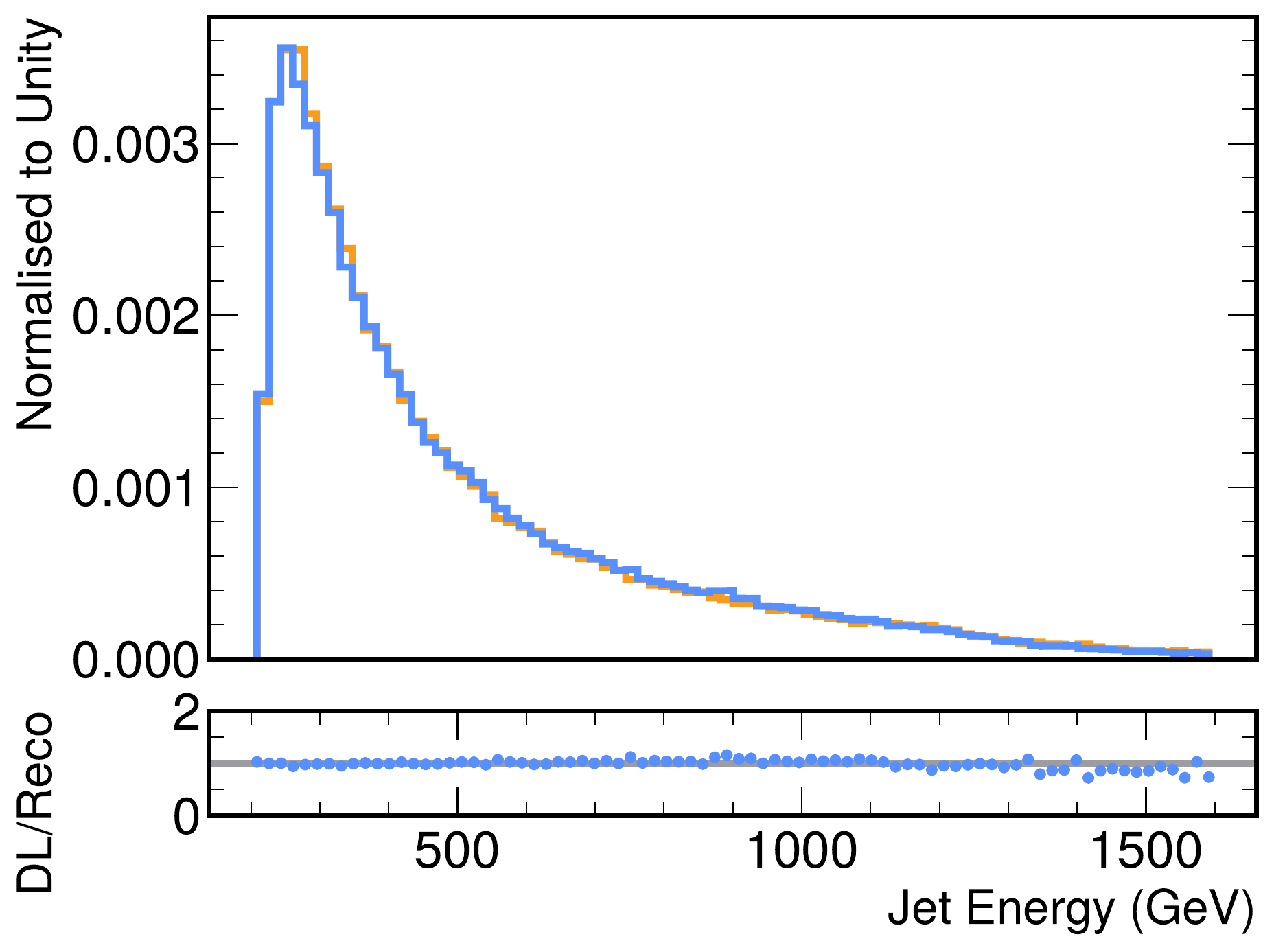}
    \includegraphics[width=0.31\textwidth]{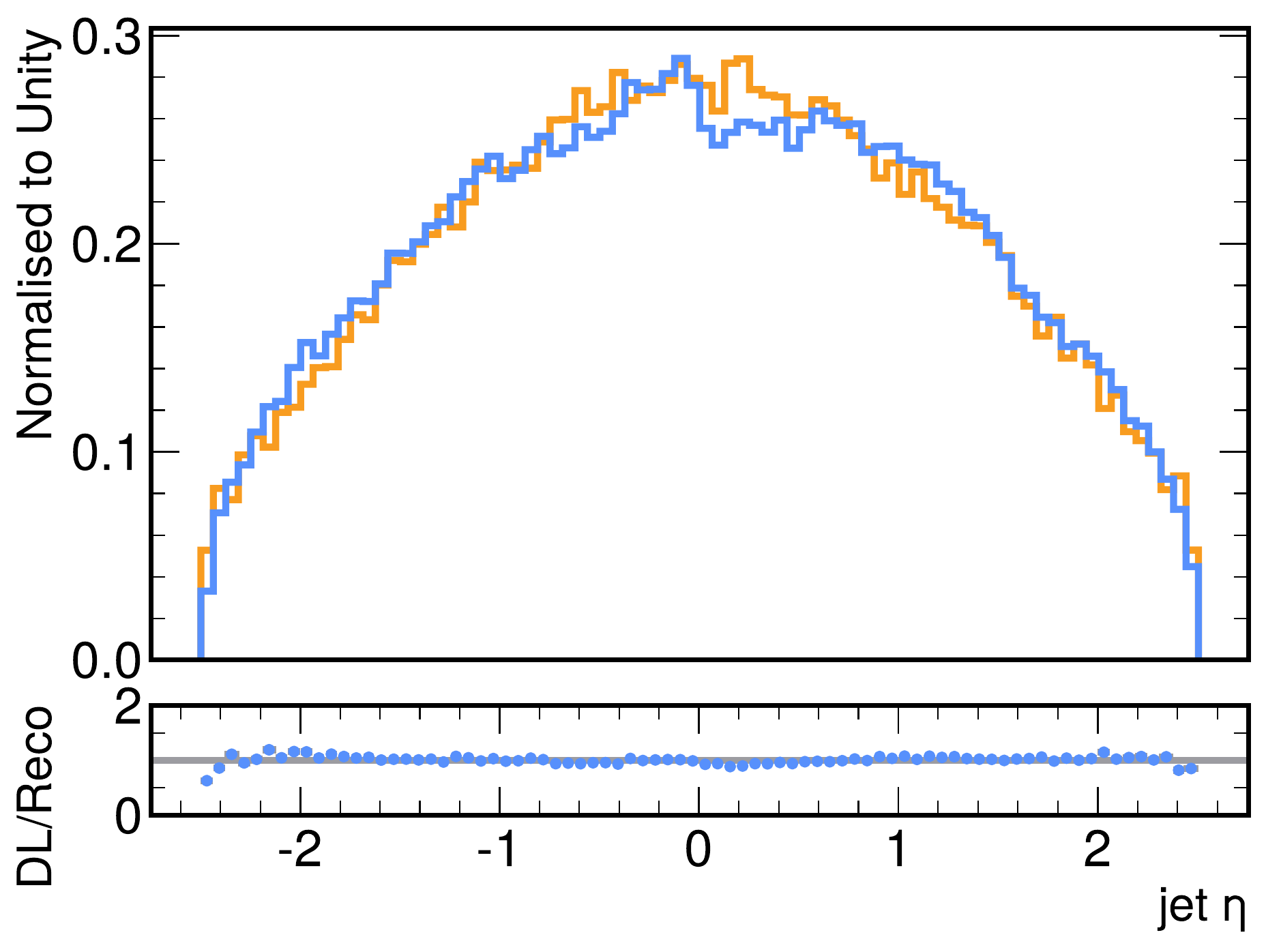}
    \includegraphics[width=0.31\textwidth]{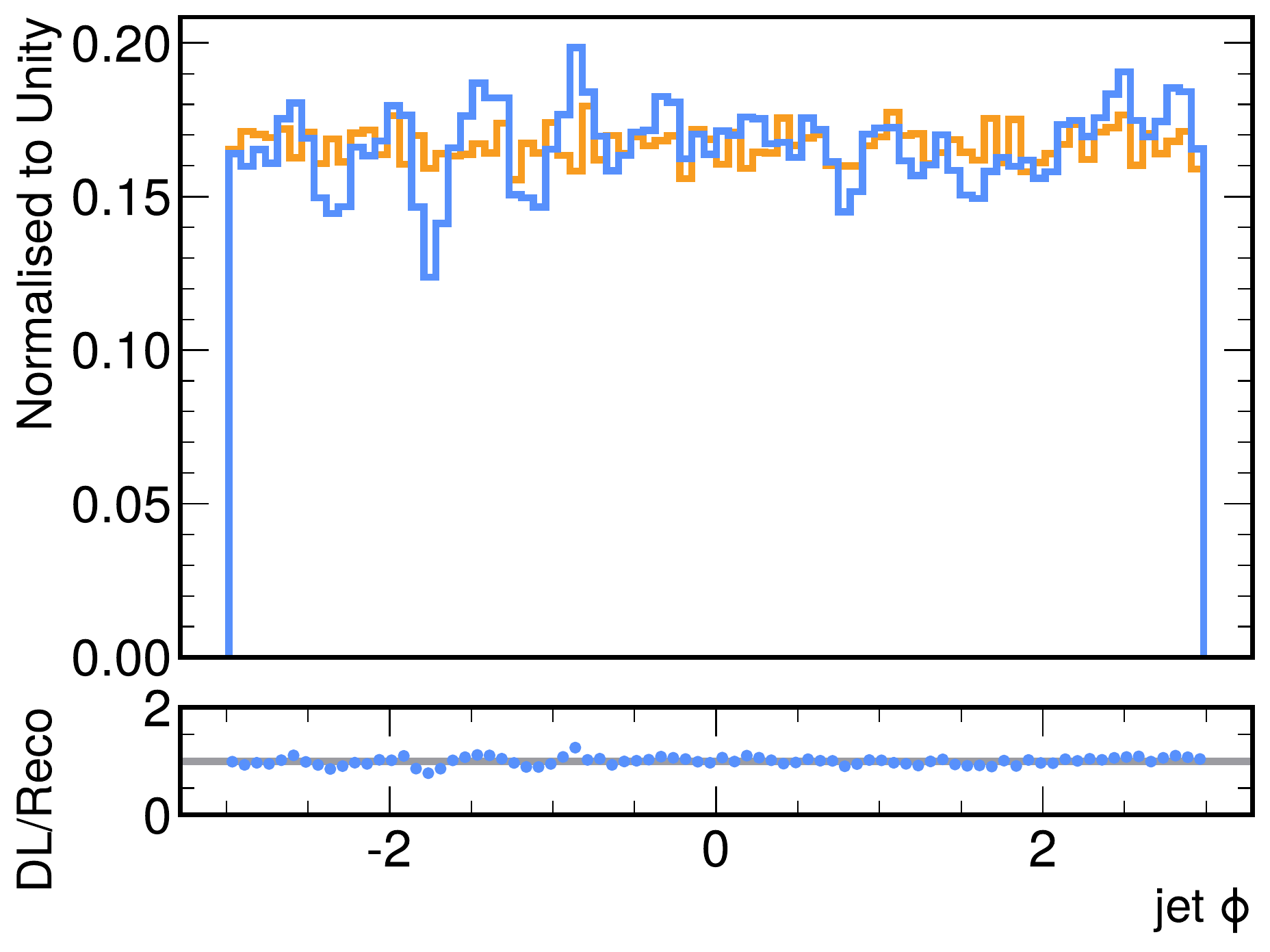}
    \caption{Comparison of the jet $p_x$ (top-left), $p_y$ (top-center), $p_z$ (top-right), energy (bottom-left), $\eta$ (bottom-center), and $\phi$ (bottom-right)  distributions, for the output (DL prediction) and target (Reconstruction) datasets. In the bottom panels, the ratio between the two distributions is shown. These distributions are obtained removing the zero-padding particles from the target list, and enforcing on the output of the DL model the same acceptance requirements that define the jet constituents (see section~\ref{sec:data}). \label{fig:jet_reco_more}}
\end{figure*}

The output list of particles is then used to analytically compute the jet kinematic properties. 
Figure~\ref{fig:jet_reco} (Fig.~\ref{fig:jet_reco_more}) shows the distribution of the jet kinematic properties explicitly used (not used) in the likelihood. The jet acceptance thresholds (see section~\ref{sec:application}) are imposed on the jet $p_T$ and $\eta$ for both the target and output jets. In general, a good agreement is observed. The residual discrepancies between the model prediction and the target reconstruction are smaller than the modelling differences typically observed between the jets data and MC reconstruction. Remarkably, once forced to learn the jet mass and transverse momentum components, the model learns to model the entire jet kinematic, including non-linear functions of the three quantities above. This aspect proves that the training process converges to a solution that preserves the main physics of the jet shower. At this stage, such a generator would be useful to generate events for most of the physics studies performed at the LHC.

In Appendix~\ref{app:nocut_plots}, we show the distribution of the jet features in the entire generation phase space, i.e., without enforcing the jet $p_T>200$~GeV requirement on the target and output jets. There, the problem of modeling the sharp $p_T$ threshold is visible. Remarkably, this issue has little impact on the agreement observed in the jet mass distribution.

\begin{figure*}
    \centering
    \includegraphics[width=0.31\textwidth]{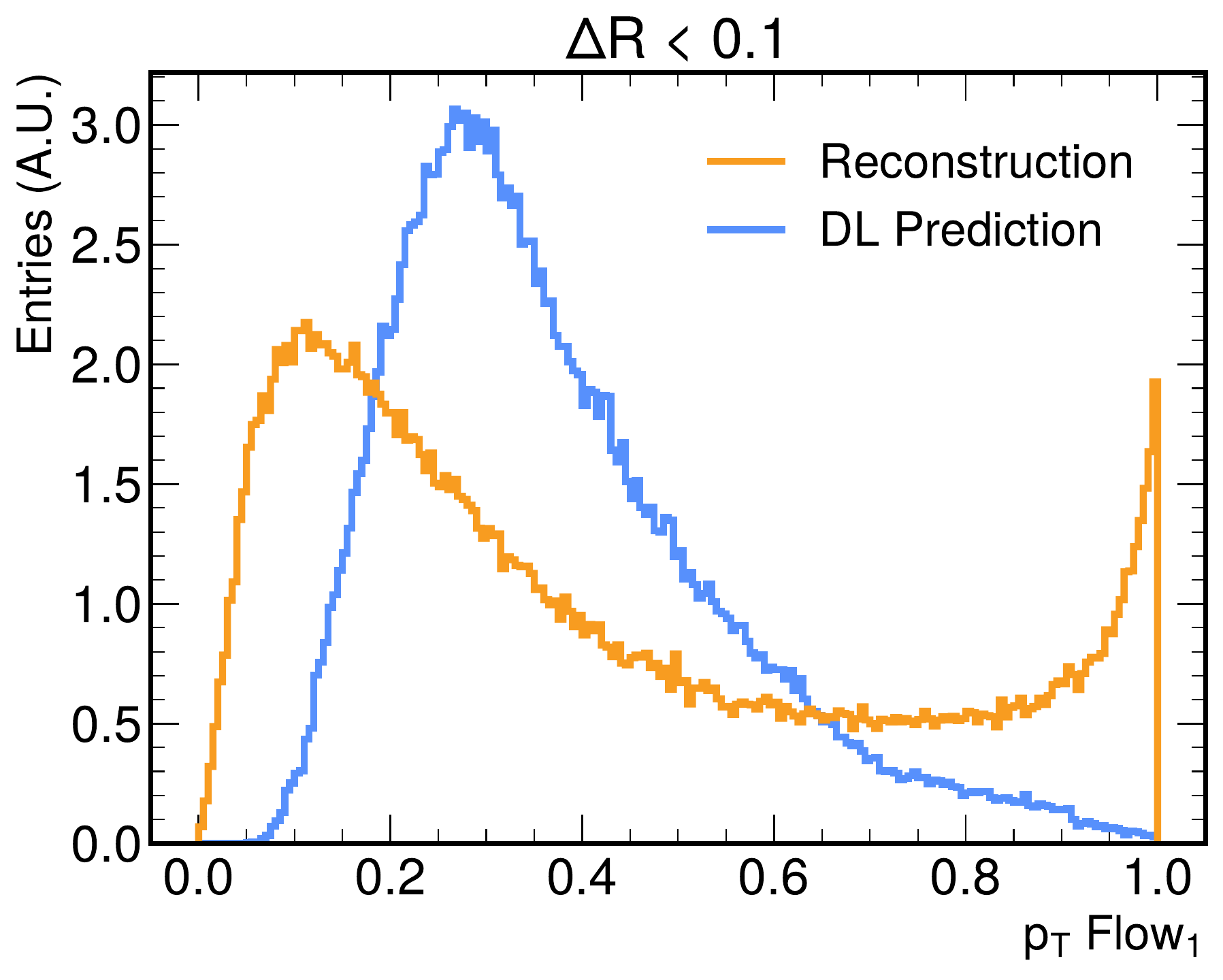}
    \includegraphics[width=0.31\textwidth]{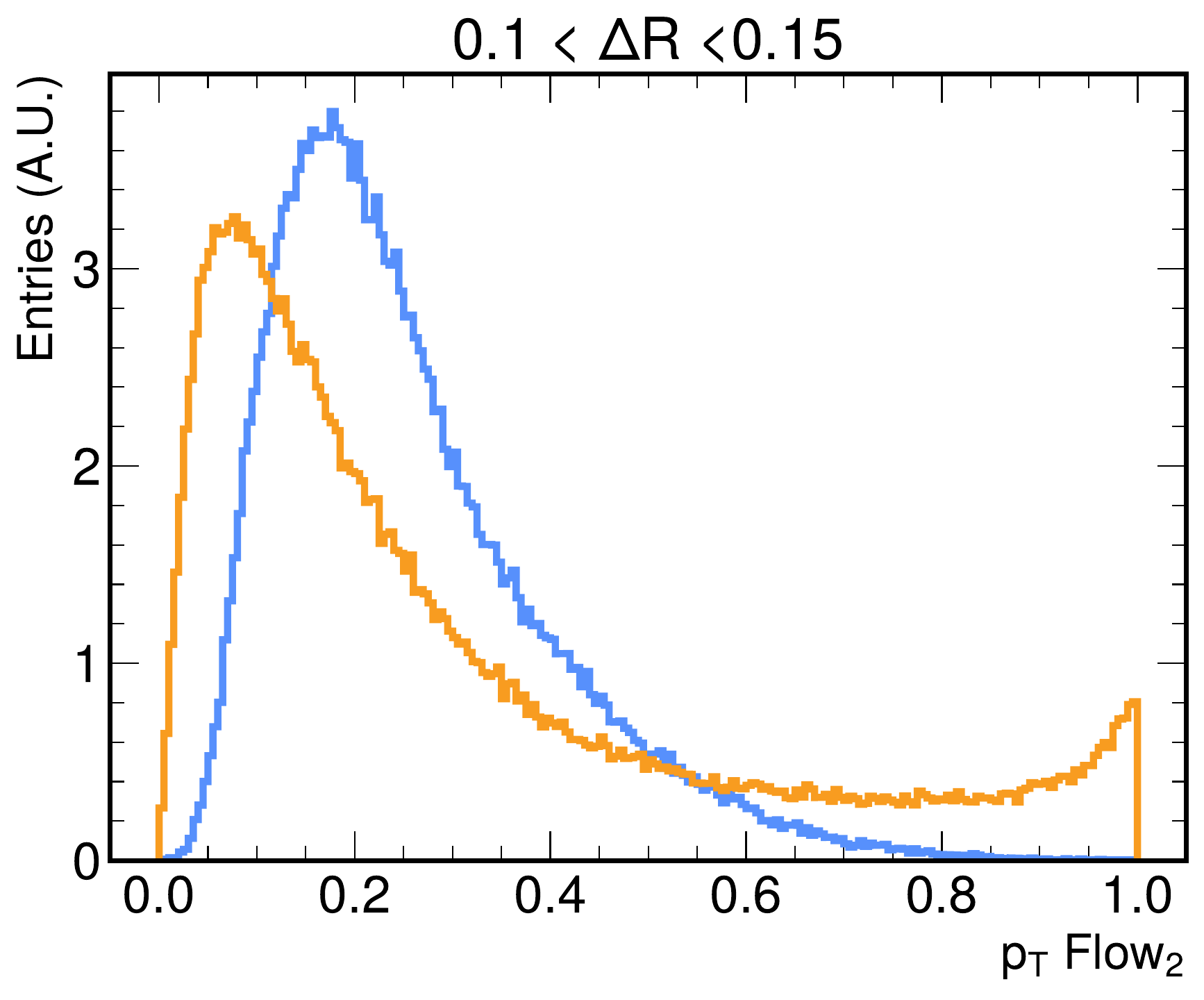}\\
    \includegraphics[width=0.31\textwidth]{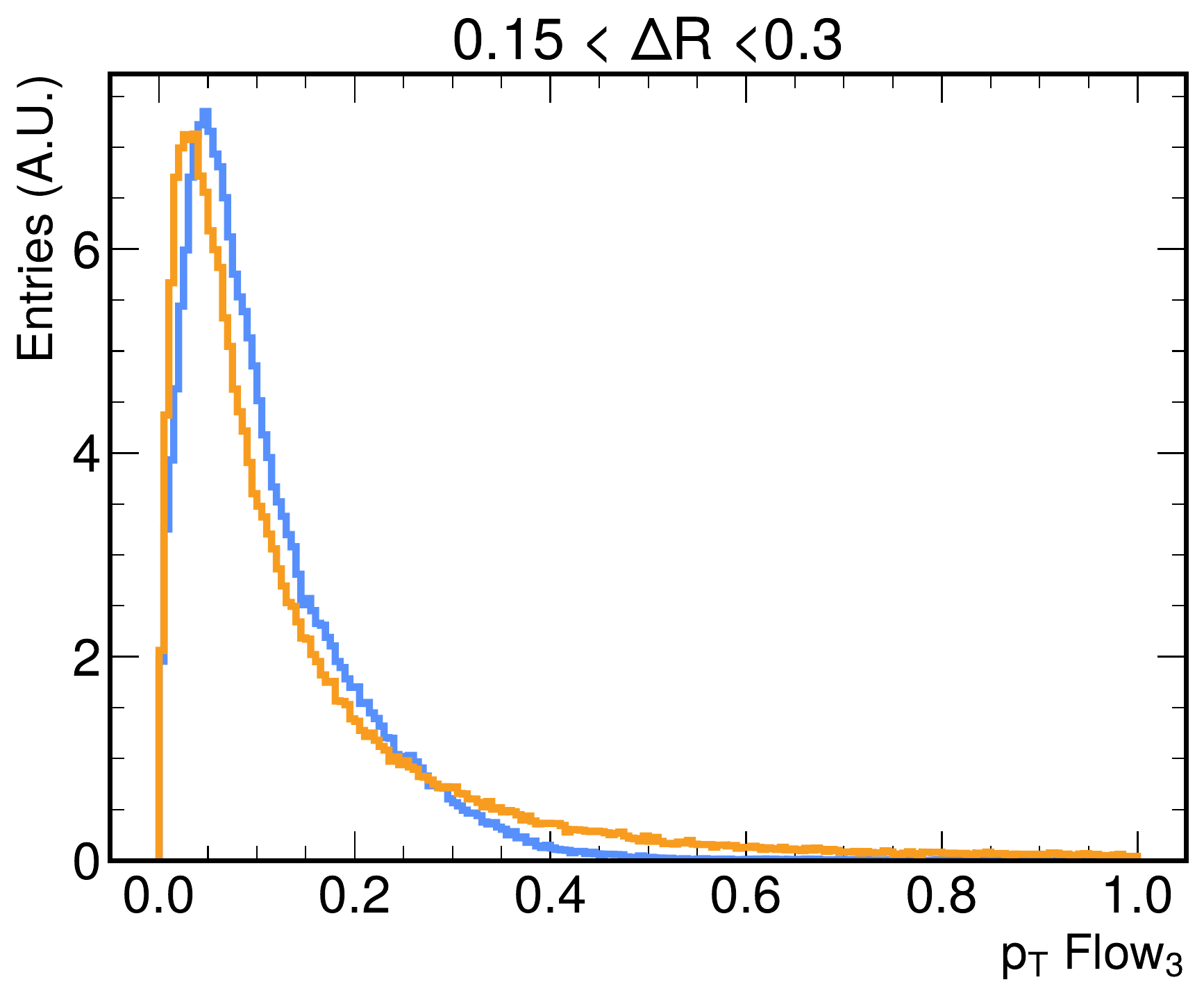}
    \includegraphics[width=0.31\textwidth]{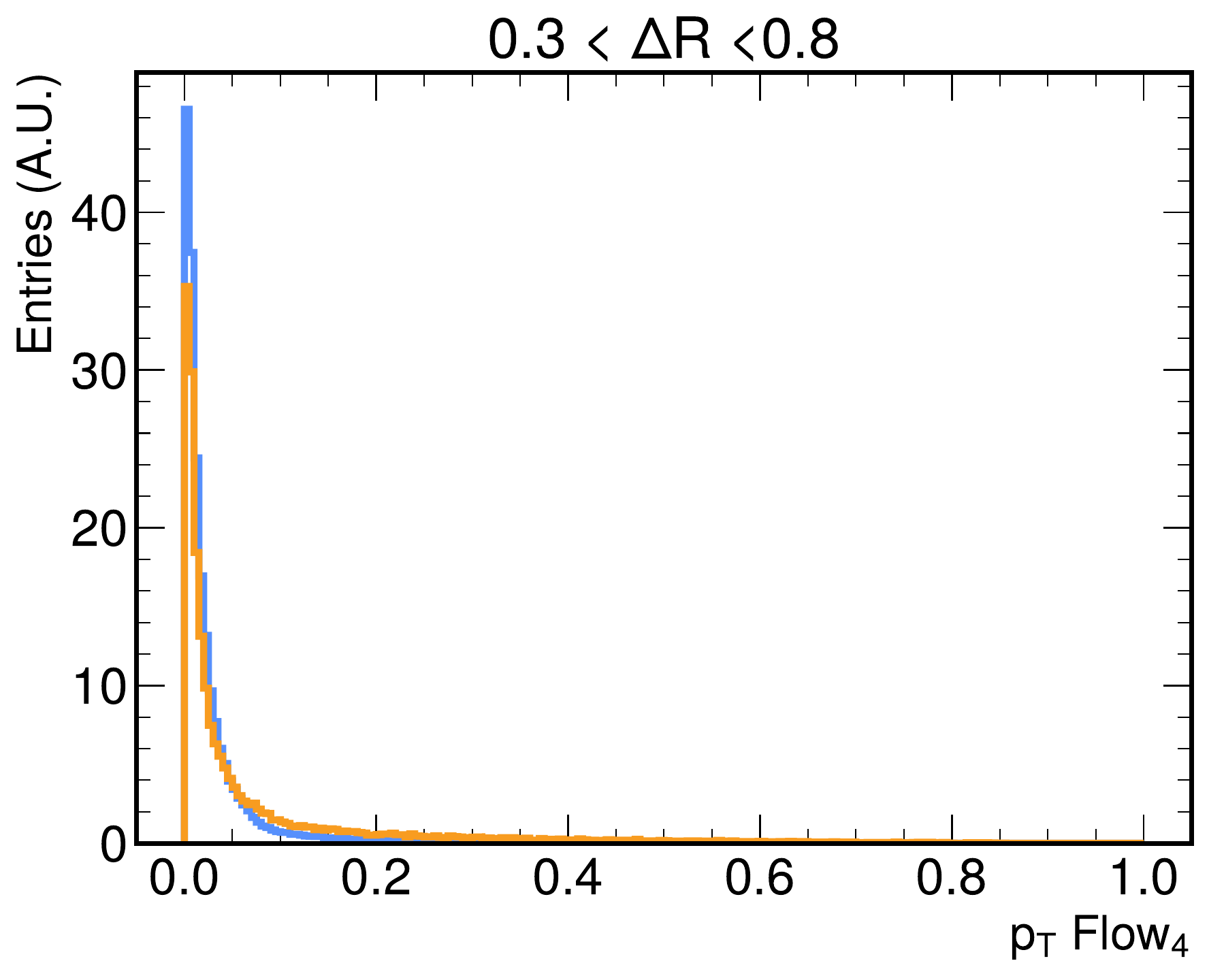}
    \caption{Distribution of the four $Flow_{n}$ quantities (see text) for the output and target jets.
    \label{fig:flows}}
\end{figure*}

While our algorithm could serve the bulk of data analyses at the LHC, it still fails in faithfully describing the jet dynamic at constituents level. In fact, we verified that jet substructure quantities are not well reproduced. This is shown in Fig.~\ref{fig:flows}, where the distribution of four momentum flows~\cite{mangano2016physics} are shown. The momentum flows are computed as $Flow_{n} = \sum_p \frac{p_T^p}{p_T^{Jet}}$, where the sum runs over all particles with distance $\Delta R = \sqrt{\Delta \phi^2 + \Delta \eta^2}$ from the jet axis falling within $(n-1)/4\times R$ and $n/4\times R$, where $R$ is the jet size parameter. We tracked the cause of the mismodeling to the noise induced by zero-momentum fake particles (both in input and target), resulting from zero-padding the jet representation to a fixed dimension. This problem could be solved moving to a graph-based VAE architecture, as in Ref.~\cite{hariri2021graph}. Through a PyTorch Geometric~\cite{fey2019fast} implementation, for example, one could avoid the need of zero-padding the datasets, possibly leading to a better representation of the jet substructure. This approach will be investigated in future studies.

The inference speed of the algorithm was measured running it on 1000 generator-level jets, and measuring the execution time. The test was performed calling the Pytorch library from a python script and running the algorithm on different hardware platforms. We obtain an inference time of 0.007 (0.004) seconds per event when running on a Intel Xeon Silver 4216 CPU (NVIDIA T4), while the traditional approaches typically require O(100) seconds per event of CPU time. This demonstrates how the proposed strategy represents a major speedup with respect to currently employed simulation algorithms. One should also keep in mind that this is an overestimate of the actual inference time in real world C++ computing environment, where tools such as ONNX run time~\cite{onnxruntime} typically offer a further speed up with respect to a python environment. 
When running on a GPU, one could further increase the throughput by running the difference on a batch of all jets in an event. 

\section{Conclusions}
\label{sec:conclusions}

We present a jet fast-simulation algorithm based on a Variational Autoencoder, trained to learn the detector response function to a generator-level jet, represented as a list of particle momenta, and returning a list of reconstructed particle momenta. This algorithm correctly captures the reconstructed jet kinematic with high accuracy. 

By bypassing the detector simulation and particle reconstruction step, an algorithm of this kind could be important to make simulation on demand a concrete possibility at the High-Luminosity LHC. 

The main strength of the current algorithm is in its speed and high accuracy when modeling jet kinematic quantities, which makes it applicable to the majority of LHC physics studies. Its main limitation stands with the poor description of the jet substructure, a consequence of the noise induced by zero-momentum ghost particles introduced to equalize the length of the input particle list. A possible solution to this problem could be the use of a graph architecture with variable-length input, which we aim at investigating in the future. 

\section*{Acknowledgement}

This work was supported by the European Research Council (ERC) under the European Union's Horizon 2020 research and innovation program (grant agreement No. 772369).
R.~K. was partially supported by an IRIS-HEP fellowship through the U.S. National Science Foundation under Cooperative Agreement OAC-1836650, and by the LHC Physics Center at Fermi National Accelerator Laboratory, managed and operated by Fermi Research Alliance, LLC under Contract No. DE-AC02-07CH11359 with the U.S. Department of Energy (DOE).
J.~D. is supported by the U.S. Department of Energy (DOE), Office of Science, Office of High Energy Physics Early Career Research program under Award No. DE-SC0021187. 
D.~G. is partially supported by the EU ICT-48 2020 project TAILOR (No. 952215).
J-R.~V. is partially supported by the European Research Council (ERC) under the European Union's Horizon 2020 research and innovation program (Grant Agreement No. 772369) and by the U.S. DOE, Office of Science, Office of High Energy Physics under Award No. DE-SC0011925, DE-SC0019227, and DE-AC02-07CH11359.
B.~O. and T.~T. are supported by grant \#2018/25225-9, SãoPaulo Research Foundation (FAPESP). B.~O. is also supported by grant \#2020/06600-3, São Paulo Research Foundation (FAPESP).
This work was supported in part by NSF awards CNS-1730158, ACI-1540112, ACI-1541349, OAC-1826967, the University of California Office of the President, and the University of California San Diego's California Institute for Telecommunications and Information Technology/Qualcomm Institute. 
Thanks to CENIC for the 100~Gpbs networks.

\cleardoublepage

\appendix
\onecolumn

\section{Appendix}
\label{app:nocut_plots}
\begin{figure}[h!]
    \centering
    \includegraphics[width=0.4\textwidth]{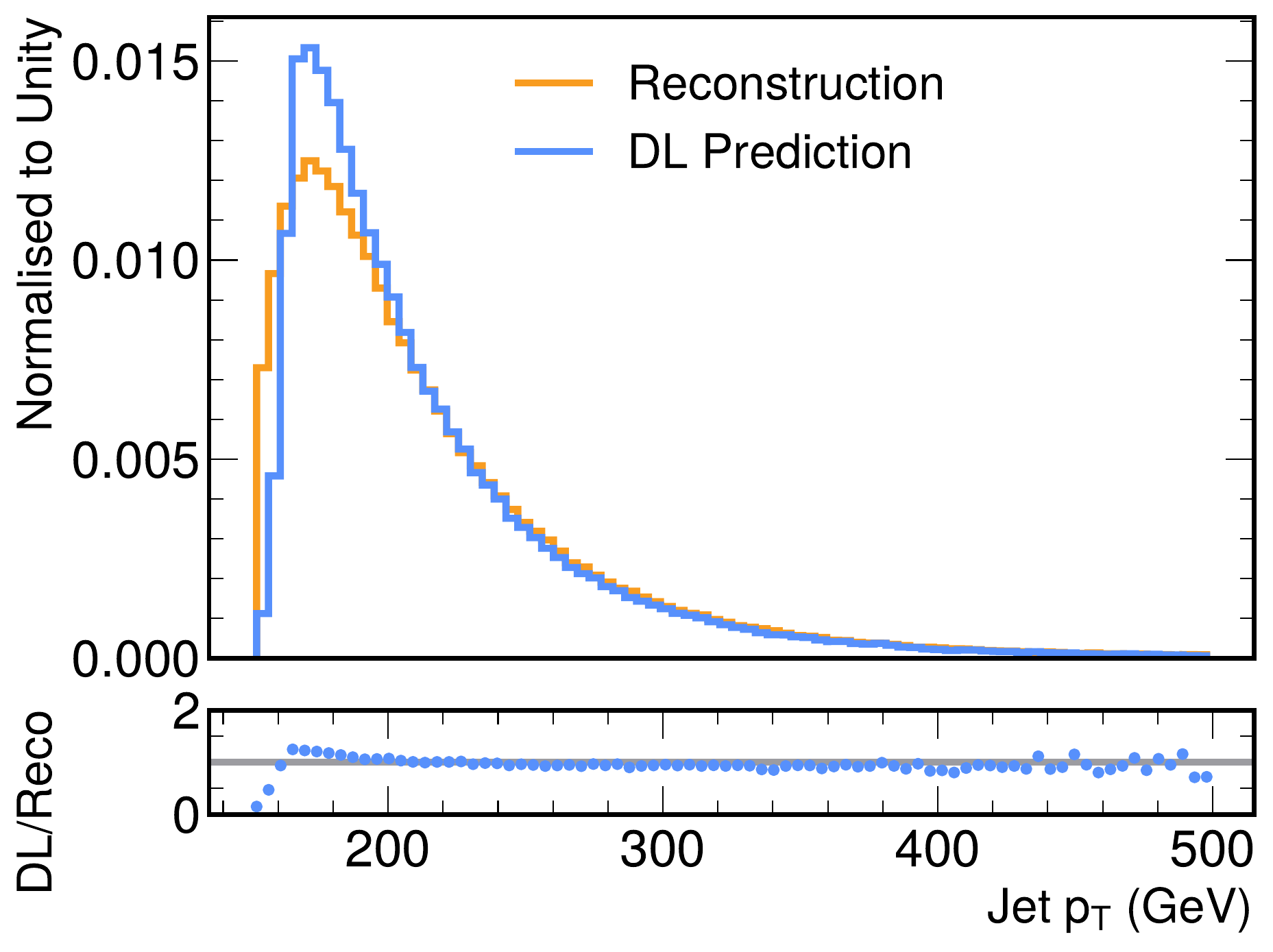}
    \includegraphics[width=0.4\textwidth]{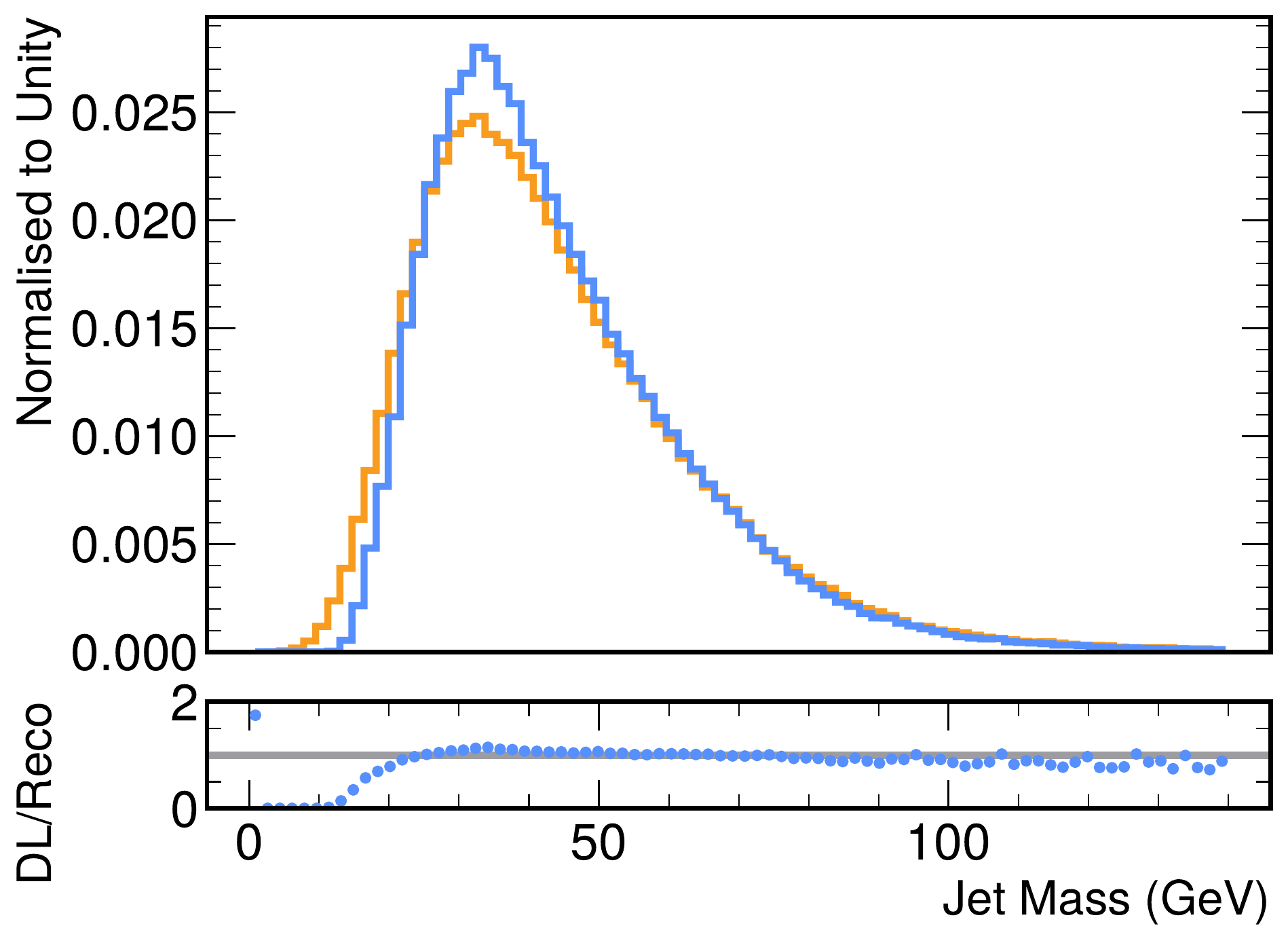}
    \caption{Comparison of the jet $p_T$ (left) and mass (right) distributions, for the output (DL prediction) and target (Reconstruction) datasets before applying the jet $p_T$ selection $p_T>200$~GeV. In the bottom panels, the ratio between the two distributions is shown. These distributions are obtained removing the zero-padding particles from the target list, and enforcing on the output of the DL model the same acceptance requirements that define the jet constituents (see section~\ref{sec:data}). \label{fig:appendix_jet_reco}} 
\end{figure}

\begin{figure}[htb!]
    \centering
    \includegraphics[width=0.31\textwidth]{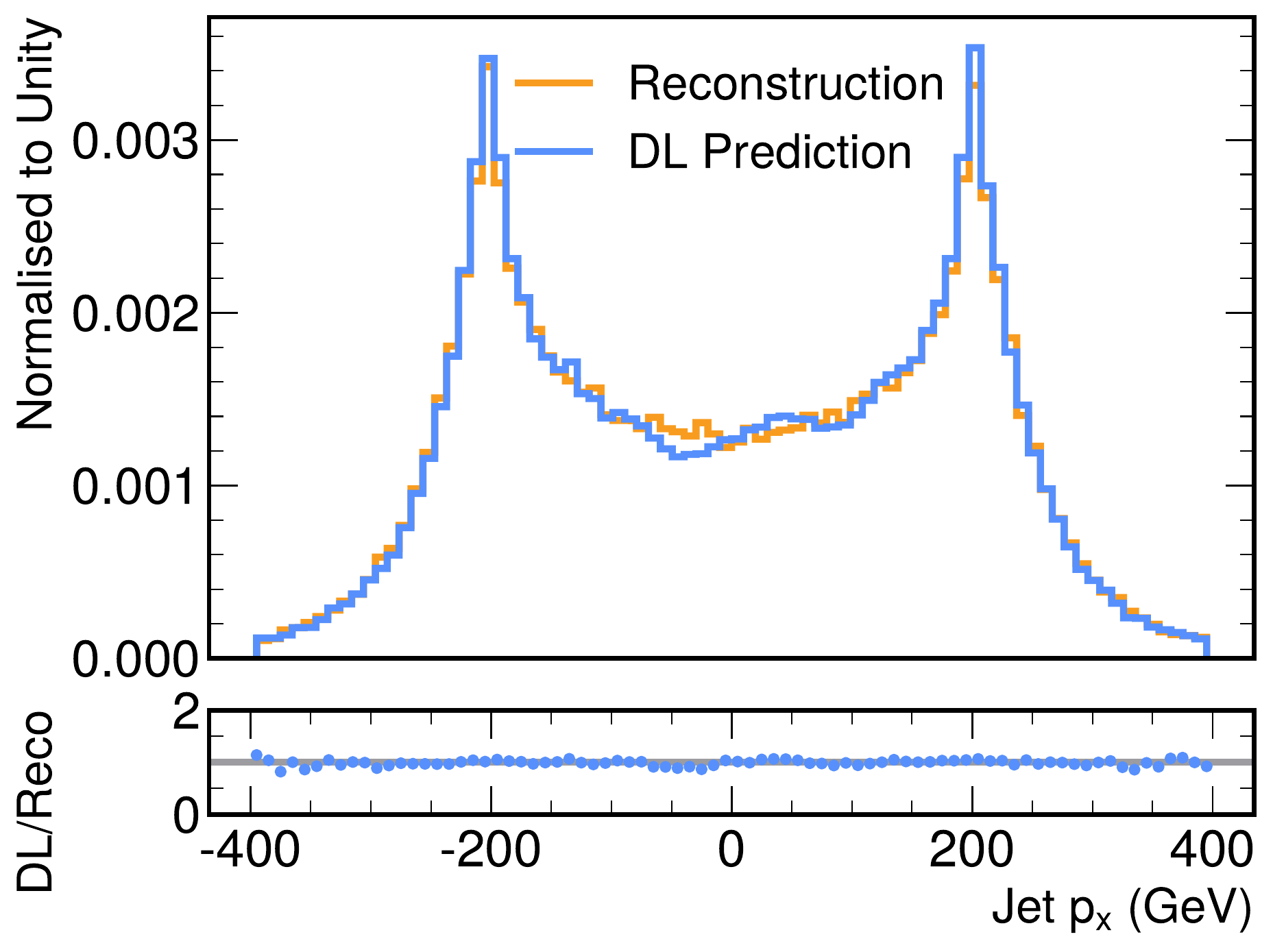} 
    \includegraphics[width=0.31\textwidth]{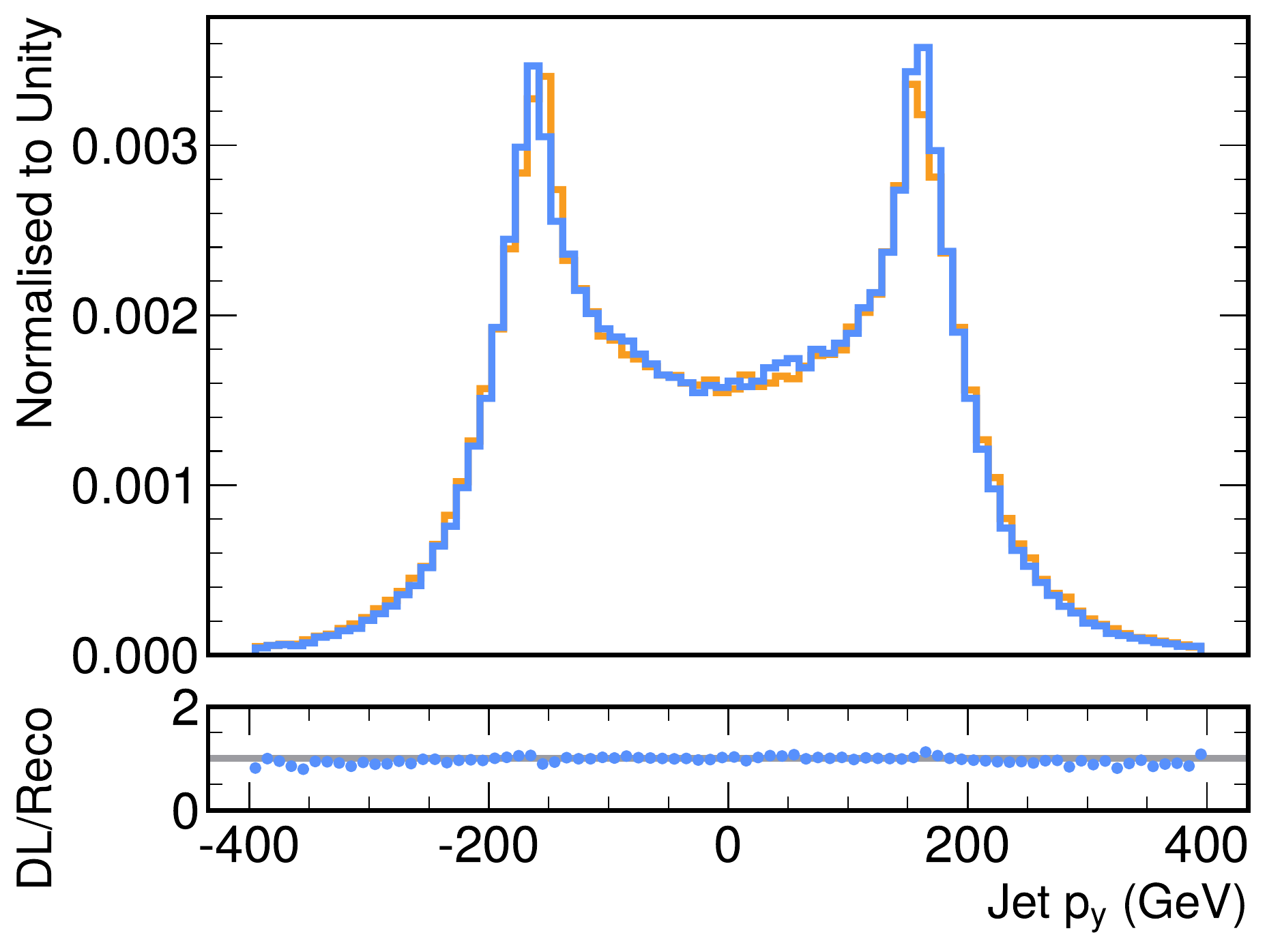} 
    \includegraphics[width=0.31\textwidth]{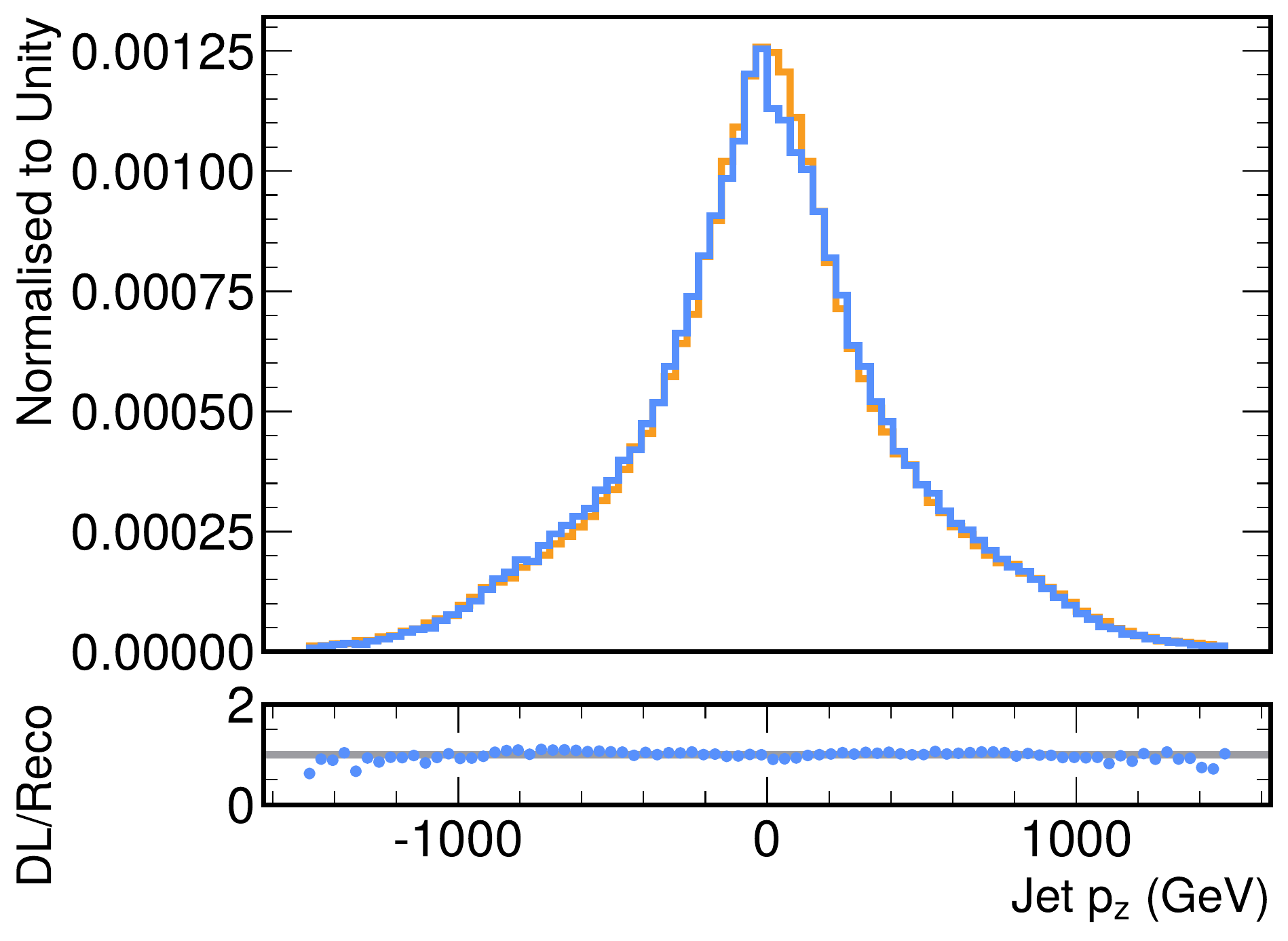} \\
    \includegraphics[width=0.31\textwidth]{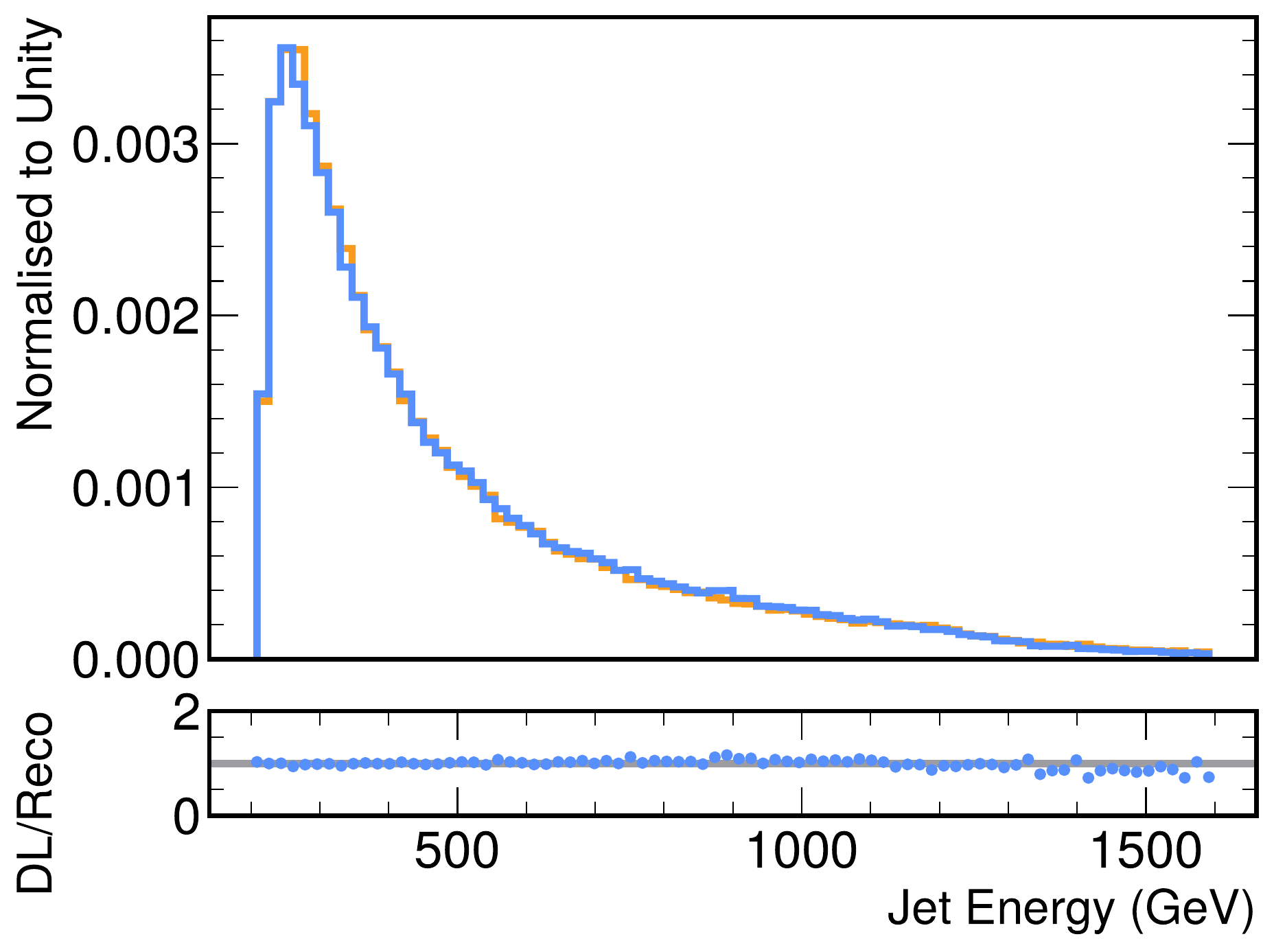}
    \includegraphics[width=0.31\textwidth]{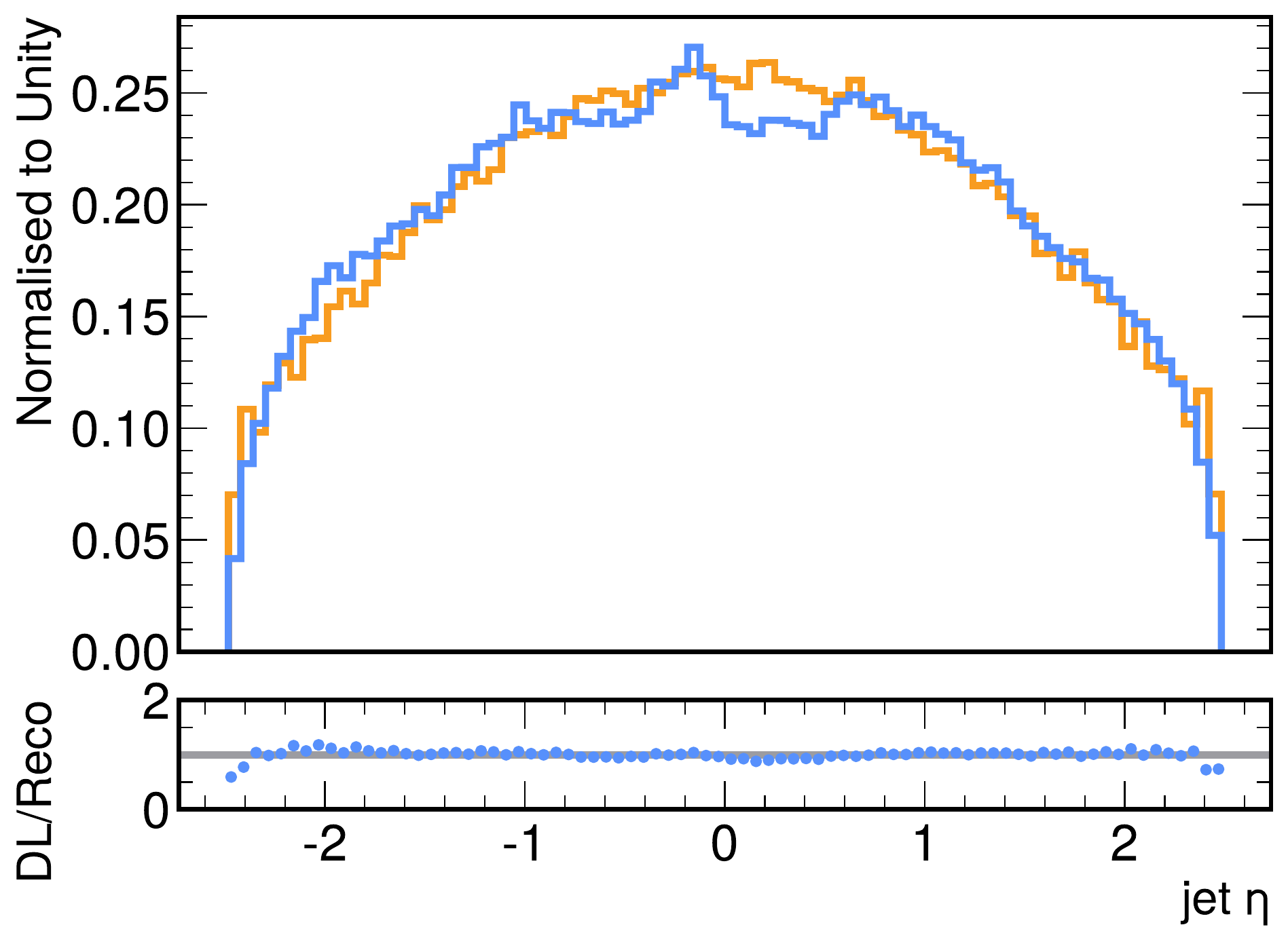}
    \includegraphics[width=0.31\textwidth]{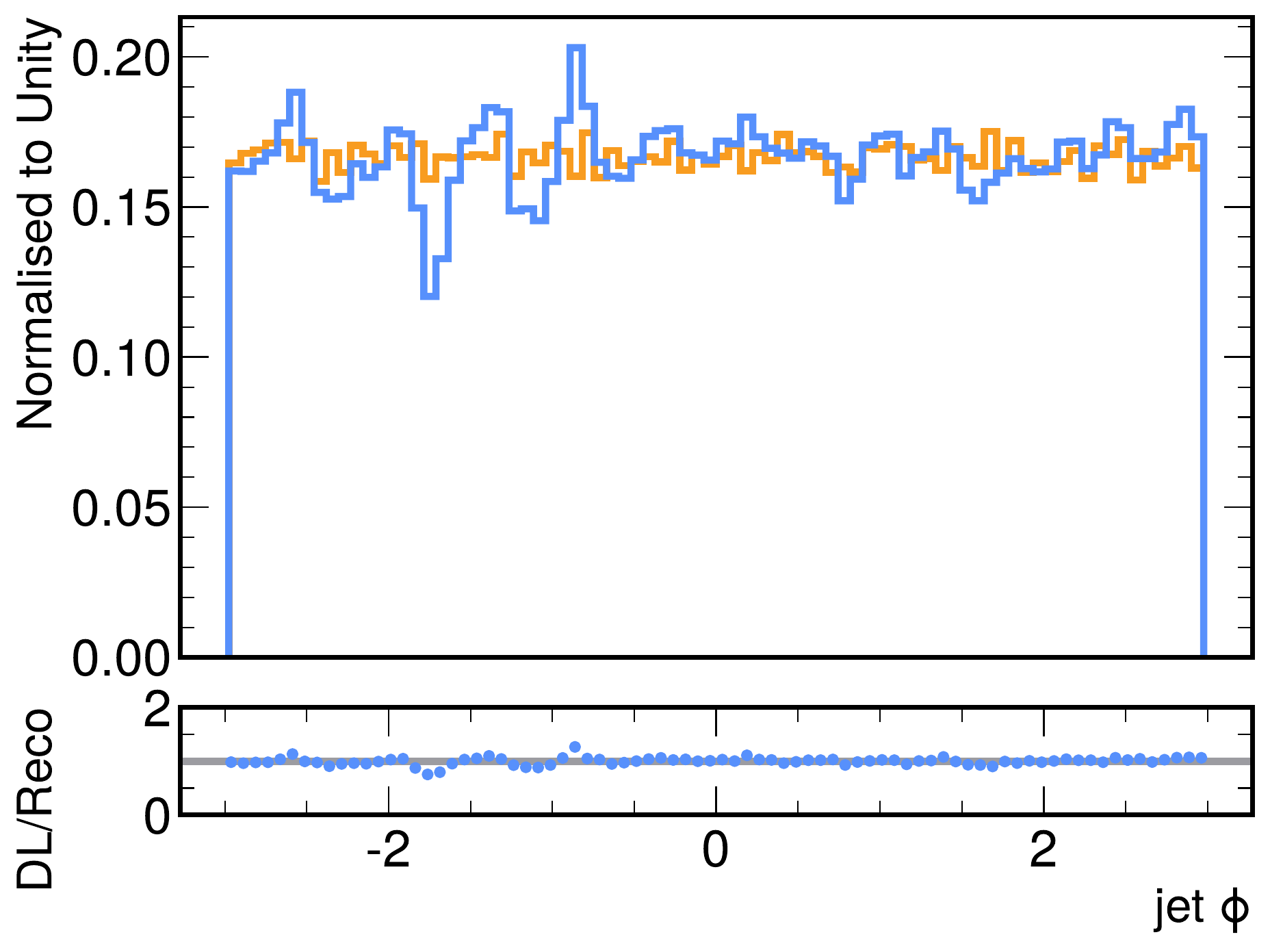}
    \caption{Comparison of the jet $p_x$ (top-left), $p_y$ (top-center), $p_z$ (top-right), energy (bottom-left), $\eta$ (bottom-center), and $\phi$ (bottom-right)  distributions, for the output (DL prediction) and target (Reconstruction) datasets before applying the jet $p_T$ selection $p_T>200$~GeV. In the bottom panels, the ratio between the two distributions is shown. These distributions are obtained removing the zero-padding particles from the target list, and enforcing on the output of the DL model the same acceptance requirements that define the jet constituents (see section~\ref{sec:data}). \label{fig:appendix_jet_reco_more}}
\end{figure}

In this appendix, we show the distribution of the jet features in the entire generation phase space before applying the jet $p_T$ selection $p_T>200$~GeV. Figure~\ref{fig:appendix_jet_reco} (Fig.~\ref{fig:appendix_jet_reco_more}) shows the distribution of the jet kinematic properties explicitly used (not used) in the likelihood within the extended jet phase space before any selections. 

\twocolumn

\bibliographystyle{unsrt}
\bibliography{bibliography}

\begin{thebibliography}{10}

\bibitem{antikt}
Matteo Cacciari, Gavin~P. Salam, and Gregory Soyez.
\newblock {The anti-$k_t$ jet clustering algorithm}.
\newblock {\em JHEP}, 04:063, 2008.

\bibitem{PFcms}
A.~M. Sirunyan et~al.
\newblock {Particle-flow reconstruction and global event description with the
  CMS detector}.
\newblock {\em JINST}, 12(10):P10003, 2017.

\bibitem{PFatlas}
Morad Aaboud et~al.
\newblock {Jet reconstruction and performance using particle flow with the
  ATLAS Detector}.
\newblock {\em Eur. Phys. J. C}, 77(7):466, 2017.

\bibitem{Agostinelli:2002hh}
S.~Agostinelli et~al.
\newblock {GEANT4: A Simulation toolkit}.
\newblock {\em Nucl. Instrum. Meth.}, A506, 2003.

\bibitem{Alves:2017she}
Johannes Albrecht et~al.
\newblock {A Roadmap for HEP Software and Computing R\&D for the 2020s}.
\newblock {\em Comput. Softw. Big Sci.}, 3(1):7, 2019.

\bibitem{Hagiwara_2013}
K.~Hagiwara, J.~Kanzaki, Q.~Li, N.~Okamura, and T.~Stelzer.
\newblock {Fast computation of MadGraph amplitudes on graphics processing unit
  (GPU)}.
\newblock {\em Eur. Phys. J. C}, 73:2608, 2013.

\bibitem{Shlomi:2020gdn}
J.~Shlomi, P.~Battaglia, and J.~R. Vlimant.
\newblock {Graph Neural Networks in Particle Physics}.
\newblock {\em Machine Learning for Science and Technology}, 2(2):021001, 2021.

\bibitem{kansal2021graph}
R.~Kansal et~al.
\newblock {Graph Generative Adversarial Networks for Sparse Data Generation in
  High Energy Physics}.
\newblock In {\em {34th Conference on Neural Information Processing Systems}},
  2020.

\bibitem{hariri2021graph}
A.~Hariri, D.~Dyachkova, and S.~Gleyzer.
\newblock Graph generative models for fast detector simulations in high energy
  physics, 2021.

\bibitem{deFavereau:2013fsa}
J.~de~Favereau et~al.
\newblock {DELPHES 3, A modular framework for fast simulation of a generic
  collider experiment}.
\newblock {\em JHEP}, 02:057, 2014.

\bibitem{sekmen2017recent}
S.~Sekmen.
\newblock {Recent Developments in CMS Fast Simulation}, 2017.

\bibitem{atlascollaboration2021atlfast3}
Georges Aad et~al.
\newblock {AtlFast3: the next generation of fast simulation in ATLAS}, 9 2021.

\bibitem{Paganini:2017hrr}
M.~Paganini, L.~de~Oliveira, and B.~Nachman.
\newblock {Accelerating Science with Generative Adversarial Networks: An
  Application to 3D Particle Showers in Multilayer Calorimeters}.
\newblock {\em Phys. Rev. Lett.}, 120(4), 2018.

\bibitem{Paganini:2017dwg}
M.~Paganini, L.~de~Oliveira, and B.~Nachman.
\newblock {CaloGAN : Simulating 3D high energy particle showers in multilayer
  electromagnetic calorimeters with generative adversarial networks}.
\newblock {\em Phys. Rev. D}, 97(1), 2018.

\bibitem{Erdmann:2018jxd}
M.~Erdmann, J.~Glombitza, and T.~Quast.
\newblock {Precise simulation of electromagnetic calorimeter showers using a
  Wasserstein Generative Adversarial Network}.
\newblock {\em Comput. Softw. Big Sci.}, 3(1), 2019.

\bibitem{Salamani:2645142}
D.~{Salamani} et~al.
\newblock Deep generative models for fast shower simulation in atlas.
\newblock In {\em 2018 IEEE 14th International Conference on e-Science
  (e-Science)}, page 348, 2018.

\bibitem{Belayneh:2019vyx}
Dawit Belayneh et~al.
\newblock {Calorimetry with deep learning: particle simulation and
  reconstruction for collider physics}.
\newblock {\em Eur. Phys. J. C}, 80(7):688, 2020.

\bibitem{Buhmann:2020pmy}
E.~Buhmann et~al.
\newblock {Getting High: High Fidelity Simulation of High Granularity
  Calorimeters with High Speed}.
\newblock {\em Comput. Softw. Big Sci.}, 5(1):13, 2021.

\bibitem{Buhmann:2021vlp}
E.~Buhmann et~al.
\newblock {Fast and Accurate Electromagnetic and Hadronic Showers from
  Generative Models}.
\newblock {\em EPJ Web Conf.}, 251:03049, 2021.

\bibitem{deOliveira:2017pjk}
L.~de~Oliveira, M.~Paganini, and B.~Nachman.
\newblock {Learning Particle Physics by Example: Location-Aware Generative
  Adversarial Networks for Physics Synthesis}.
\newblock {\em Comput. Softw. Big Sci.}, 1(1), 2017.

\bibitem{Musella:2018rdi}
P.~Musella and F.~Pandolfi.
\newblock {Fast and Accurate Simulation of Particle Detectors Using Generative
  Adversarial Networks}.
\newblock {\em Comput. Softw. Big Sci.}, 2(1), 2018.

\bibitem{Carrazza:2019cnt}
S.~Carrazza and F.~A. Dreyer.
\newblock {Lund jet images from generative and cycle-consistent adversarial
  networks}.
\newblock {\em Eur. Phys. J. C}, 79(11), 2019.

\bibitem{Otten:2019hhl}
Sydney Otten et~al.
\newblock {Event Generation and Statistical Sampling for Physics with Deep
  Generative Models and a Density Information Buffer}.
\newblock {\em Nature Commun.}, 12(1):2985, 2021.

\bibitem{Hashemi:2019fkn}
B.~Hashemi et~al.
\newblock {LHC analysis-specific datasets with Generative Adversarial
  Networks}, 2019.

\bibitem{DiSipio:2019imz}
R.~Di~Sipio et~al.
\newblock {DijetGAN: A Generative-Adversarial Network Approach for the
  Simulation of QCD Dijet Events at the LHC}.
\newblock {\em J. High Energy Phys.}, 08, 2020.

\bibitem{Butter:2019cae}
A.~Butter, T.~Plehn, and R.~Winterhalder.
\newblock {How to GAN LHC Events}.
\newblock {\em SciPost Phys.}, 7, 2019.

\bibitem{Erdmann:2018kuh}
M.~Erdmann et~al.
\newblock {Generating and refining particle detector simulations using the
  Wasserstein distance in adversarial networks}.
\newblock {\em Comput. Softw. Big Sci.}, 2(1), 2018.

\bibitem{Goodfellow:2014upx}
I.~J. Goodfellow et~al.
\newblock {Generative Adversarial Networks}, 2014.

\bibitem{arjovsky2017wasserstein}
M.~Arjovsky, S.~Chintala, and L.~Bottou.
\newblock {Wasserstein GAN}, 2017.

\bibitem{1704.00028}
I.~Gulrajani et~al.
\newblock Improved training of {Wasserstein GANs}, 2017.

\bibitem{kingma2014auto}
D.~P {Kingma} and M.~{Welling}.
\newblock {Auto-Encoding Variational Bayes}, 2013.

\bibitem{Martinez:2019jlu}
J.~Arjona~Mart\'\i{}nez et~al.
\newblock {Particle Generative Adversarial Networks for full-event simulation
  at the LHC and their application to pileup description}.
\newblock {\em J. Phys. Conf. Ser.}, 1525(1):012081, 2020.

\bibitem{Belavin_2020}
V.~Belavin and A.~Ustyuzhanin.
\newblock Electromagnetic shower generation with graph neural networks.
\newblock {\em Journal of Physics: Conference Series}, 1525:012105, 2020.

\bibitem{Krause:2021ilc}
C.~Krause and D.~Shih.
\newblock {CaloFlow: Fast and Accurate Generation of Calorimeter Showers with
  Normalizing Flows}, 2021.

\bibitem{Lanusse}
F.~Lanusse.
\newblock Machine learning in cosmology, 2019.
\newblock ACAT 2019, Saas Fee (CH).

\bibitem{Orzari:2021suh}
B.~Orzari et~al.
\newblock {Sparse Data Generation for Particle-Based Simulation of Hadronic
  Jets in the LHC}.
\newblock In {\em {38th International Conference on Machine Learning
  Conference}}, 2021.

\bibitem{Chen:2020uds}
C.~Chen et~al.
\newblock {Data Augmentation at the LHC through Analysis-specific Fast
  Simulation with Deep Learning}.
\newblock {\em Comput Softw Big Sci}, 5(15), 2021.

\bibitem{Sj_strand_2015}
T.~Sjöstrand et~al.
\newblock An introduction to pythia 8.2.
\newblock {\em Computer Physics Communications}, 191, 2015.

\bibitem{touranakou_mary_2022_6047873}
M.~Touranakou et~al.
\newblock {Particle-based Fast Jet Simulation at the LHC with Variational
  Autoencoders: generator-level and reconstruction-level jets dataset}, 2022.
\newblock \url{https://doi.org/10.5281/zenodo.6047873}.

\bibitem{torch}
A.~Paszke et~al.
\newblock Pytorch: An imperative style, high-performance deep learning library,
  2019.
\newblock \url{https://arxiv.org/pdf/1912.01703.pdf}.

\bibitem{rezende2016variational}
D.~J. Rezende and S.~Mohamed.
\newblock Variational inference with normalizing flows, 2016.

\bibitem{Higgins2017betaVAELB}
I.~Higgins et~al.
\newblock $\beta$-{VAE}: Learning basic visual concepts with a constrained
  variational framework.
\newblock In {\em 5th International Conference on Learning Representations},
  2017.

\bibitem{chamfer}
H.~Fan, H.~Su, and L.~Guibas.
\newblock A point set generation network for 3d object reconstruction from a
  single image, 2016.

\bibitem{Belayneh_2020}
D.~Belayneh et~al.
\newblock Calorimetry with deep learning: particle simulation and
  reconstruction for collider physics.
\newblock {\em The European Physical Journal C}, 80(7), 2020.

\bibitem{CMS:2015xau}
V.~Khachatryan et~al.
\newblock {Search for narrow resonances decaying to dijets in proton-proton
  collisions at $\sqrt{s} =$ 13 TeV}.
\newblock {\em Phys. Rev. Lett.}, 116(7):071801, 2016.

\bibitem{ATLAS:2017eqx}
M.~Aaboud et~al.
\newblock {Search for new phenomena in dijet events using 37 fb$^{-1}$ of $pp$
  collision data collected at $\sqrt{s}=$13 TeV with the ATLAS detector}.
\newblock {\em Phys. Rev. D}, 96(5):052004, 2017.

\bibitem{CMS:2019gwf}
A.~M. Sirunyan et~al.
\newblock {Search for high mass dijet resonances with a new background
  prediction method in proton-proton collisions at $\sqrt{s} =$ 13 TeV}.
\newblock {\em JHEP}, 05:033, 2020.

\bibitem{CMS:2018ucw}
A.~M. Sirunyan et~al.
\newblock {Search for new physics in dijet angular distributions using
  proton\textendash{}proton collisions at $\sqrt{s}=$ 13 TeV and constraints on
  dark matter and other models}.
\newblock {\em Eur. Phys. J. C}, 78(9):789, 2018.

\bibitem{adam}
D.~K. Kingma and J.~Ba.
\newblock Adam: A method for stochastic optimization, 2015.

\bibitem{kullback1951information}
S.~Kullback and R.~A. Leibler.
\newblock On information and sufficiency.
\newblock {\em The annals of mathematical statistics}, 22(1):79--86, 1951.

\bibitem{mangano2016physics}
M.~L. Mangano et~al.
\newblock {Physics at a 100 TeV pp Collider: Standard Model Processes}, 2016.

\bibitem{fey2019fast}
M.~Fey and J.~E. Lenssen.
\newblock Fast graph representation learning with pytorch geometric, 2019.

\bibitem{onnxruntime}
Onnx runtime.
\newblock \url{https://www.onnxruntime.ai}, 2021.

\end{thebibliography}

\end{document}